\documentclass[conference]{IEEEtran}
\IEEEoverridecommandlockouts
\usepackage{tabularx}
\usepackage{array}
\usepackage{cite}
\usepackage{amsmath,amssymb,amsfonts,bm}
\usepackage{graphicx}
\usepackage{textcomp}
\usepackage{xcolor}
\usepackage{amsmath}
\usepackage{acronym}
\usepackage{hyperref}
\usepackage{bbm}
\usepackage{graphicx}
\usepackage{subcaption}
\usepackage{booktabs}
\usepackage{algorithm}
\usepackage{algpseudocode}
\usepackage[inline]{enumitem}

\def\BibTeX{{\rm B\kern-.05em{\sc i\kern-.025em b}\kern-.08em
    T\kern-.1667em\lower.7ex\hbox{E}\kern-.125emX}}

\newcommand{\algoname}{Algorithm}
\newcommand{\sectionname}{Sect.}
\newcommand{\sectionsname}{Sects.}

\renewcommand{\tablename}{Table}
\newcommand{\tablesname}{Tables}

\newcommand{\meter}{\,\mathrm{m}}
\newcommand{\mW}{\,\mathrm{mW}}
\newcommand{\dB}{\,\mathrm{dB}}
\newcommand{\MHz}{\,\mathrm{MHz}}
\newcommand{\GHz}{\,\mathrm{GHz}}

\newcommand{\ms}{\,\mathrm{ms}}
\newcommand{\second}{\,\mathrm{s}}
\newcommand{\Htran}{\mathsf{H}}
\newcommand{\Ttran}{\mathsf{T}}  

\newcommand{\bpsHz}{\,\mathrm{b/s/Hz}}

\usepackage{tikz}
\usetikzlibrary{arrows.meta, positioning}
\acrodef{AWGN}{additive white Gaussian noise}
\acrodef{AdamW}{adaptive moment estimation with weight decay}
\acrodef{B5G}{beyond-5G}
\acrodef{CSA}{central satellite}
\acrodef{CFmMIMO}{cell-free massive \acs{MIMO}}
\acrodef{mMIMO}{massive \acs{MIMO}}
\acrodef{DPD} {disk Poisson Process} 
\acrodef{DRA}{direct radiating antenna}
\acrodef{DC}{direct current}
\acrodef{DoD}{depth of discharge}
\acrodef{eMBB}{enhanced mobile broadband}
\acrodef{E-GPW}{extended gridded population of world database}
\acrodef{FDM}{frequency division multiplexing}
\acrodef{FIR}{finite impulse response}
\acrodef{TFoA}{thinned \acs{FoA}}
\acrodef{FoA}{formation of arrays}
\acrodef{GEO}{geostationary Earth orbit}
\acrodef{GSO}{geosynchronous Earth orbit}
\acrodef{H-RRM}{heuristic radio resource management}
\acrodef{KPI}{key performance indicator}
\acrodef{IF}{intermediate frequency}
\acrodef{LEO}{low Earth orbit}
\acrodef{LoS}{line-of-sight}
\acrodef{MAI}{multiple access interference}
\acrodef{MB}{multi beam}
\acrodef{MEO}{medium Earth orbit}
\acrodef{mMTC}{massive machine-type communications}
\acrodef{eMTC}{enhanced machine-type communications}
\acrodef{PDD}{Poisson disk distribution}
\acrodef{EIRP}{effective isotropic radiated power}
\acrodef{NTN}{non terrestrial network}
\acrodef{URLLC}{ultra-reliable low-latency communications}
\acrodef{ECEF}{Earth-centered Earth-fixed}
\acrodef{GPS}{global positioning system}
\acrodef{HTFS}{high throughput fractionated satellite}
\acrodef{UHF}{ultra-high frequency}
\acrodef{FF}{formation flying}
\acrodef{MIMO}{multiple input multiple output}
\acrodef{M-MIMO}{massive multiple input multiple output}
\acrodef{GNC}{guidance navigation and control}
\acrodef{GNSS}{global navigation satellite system}
\acrodef{OFDM}{orthogonal frequency division multiplexing}
\acrodef{PHY}{physical-layer}
\acrodef{SA}{sub-array}
\acrodef{SADM}{solar array drive mechanism}
\acrodef{SG}{solar generator}
\acrodef{SNR}{signal-to-noise ratio}
\acrodef{SIR}{signal-to-interference ratio}
\acrodef{SINR}{signal-to-interference-plus-noise ratio}
\acrodef{SSPA}{solid-state power amplifier}
\acrodef{RF}{radio frequency}
\acrodef{R-GEO}{regional \acs{GEO}}
\acrodef{UT}{user terminal}
\acrodef{UE}{user equipment}
\acrodef{QVD}{quantized virtual distancing}
\acrodef{TDM}{time division multiplexing}
\acrodef{R-GEO}{regional GEO}
\acrodef{RRM}{radio resource management}
\acrodef{SSPA}{solid-state power amplifier}
\acrodef{BLER}{block error rate}
\acrodef{DL}{downlink}
\acrodef{UL}{uplink}
\acrodef{PSD}{power spectral density}
\acrodef{UDSM}{ultra deep sub-micron}
\acrodef{PEPS}{power energy platform simulation}
\acrodef{SE}{spectral efficiency}
\acrodef{wrt}{with respect to}
\acrodef{OBP}{on-board digital processor}
\acrodef{UDSM}{ultra-deep-sub-micron}

\acrodef{3GPP}{third generation partnership project}
\acrodef{AWGN}{additive white Gaussian noise}
\acrodef{B5G}{beyond-5G}
\acrodef{CS}{central satellite}
\acrodef{DRA}{direct radiating antenna}
\acrodef{DC}{direct current}
\acrodef{DoD}{depth of discharge}
\acrodef{eMBB}{enhanced mobile broadband}
\acrodef{FDM}{frequency division multiplexing}
\acrodef{FIR}{finite impulse response}
\acrodef{TFoA}{thinned \acs{FoA}}
\acrodef{FoA}{formation of arrays}
\acrodef{GEO}{geostationary Earth orbit}
\acrodef{GSO}{geosynchronous Earth orbit}
\acrodef{KPI}{key performance indicator}
\acrodef{IF}{intermediate frequency}
\acrodef{LEO}{low Earth orbit}
\acrodef{LoS}{line-of-sight}
\acrodef{MAI}{multiple access interference}
\acrodef{MEO}{medium Earth orbit}
\acrodef{mMTC}{massive machine-type communications}
\acrodef{eMTC}{enhanced machine-type communications}
\acrodef{EIRP}{effective isotropic radiated power}
\acrodef{NTN}{non terrestrial network}
\acrodef{URLLC}{ultra-reliable low-latency communications}
\acrodef{ECEF}{Earth-centered Earth-fixed}
\acrodef{GPS}{global positioning system}
\acrodef{HTFS}{high throughput fractionated satellite}
\acrodef{HTS}{high throughput satellite}
\acrodef{UHF}{ultra-high frequency}
\acrodef{FF}{formation flying}
\acrodef{MIMO}{multiple input multiple output}
\acrodef{GNC}{guidance navigation and control}
\acrodef{NTN}{non-terrestrial network}
\acrodef{GNSS}{global navigation satellite system}
\acrodef{OFDM}{orthogonal frequency division multiplexing}
\acrodef{PHY}{physical-layer}
\acrodef{SA}{satellite array}
\acrodef{SADM}{solar array drive mechanism}
\acrodef{SG}{solar generator}
\acrodef{SNR}{signal-to-noise ratio}
\acrodef{SIR}{signal-to-interference ratio}
\acrodef{SINR}{signal-to-interference-plus-noise ratio}
\acrodef{SSPA}{solid-state power amplifier}
\acrodef{RF}{radio frequency}
\acrodef{R-GEO}{regional \acs{GEO}}
\acrodef{UT}{user terminal}
\acrodef{UE}{user equipment}
\acrodef{TDM}{time division multiplexing}
\acrodef{RRM}{radio resource management}
\acrodef{SSPA}{solid-state power amplifier}
\acrodef{BLER}{block error rate}
\acrodef{DL}{down-link}
\acrodef{UL}{up-link}
\acrodef{RV}{random variable}
\acrodef{PSD}{power spectral density}
\acrodef{UDSM}{ultra deep sub-micron}
\acrodef{PEPS}{power energy platform simulation}
\acrodef{SE}{spectral efficiency}
\acrodef{wrt}{with respect to}
\acrodef{OBP}{on-board digital processor}
\acrodef{UDSM}{ultra-deep-sub-micron}
\acrodef{UC}{user-centric}
\acrodef{CSI}{channel state information}
\acrodef{PAC}{per-antenna constraint}
\acrodef{FPAC}{fair \acs{PAC}}
\acrodef{MPC}{maximum power constraint}
\acrodef{AWGN}{additive white Gaussian noise}
\acrodef{TX}{transmit}
\acrodef{MMSE}{minimum mean square error}
\acrodef{MF}{matched filter}
\acrodef{ZF}{zero forcing}
\acrodef{ELSA}{enhanced logarithmic spiral array}
\acrodef{UPA}{uniform planar array}
\acrodef{NB}{narrowband}
\acrodef{WB}{wideband}
\acrodef{BFN}{beamforming network}
\acrodef{MD-MIQP}{minimum distance mixed integer quadratic problem}
\acrodef{PDF}{probability density function}
\acrodef{NPR}{noise-to-power ratio}
\acrodef{HPA}{high power amplifier}
\acrodef{CF}{cell-free}
\acrodef{3GPP}{third generation partnership project}
\acrodef{UC-MIMO}{user centric MIMO}
\acrodef{VHTS}{very high throughput satellites}
\acrodef{PM-MIMO}{pragmatic M-MIMO}
\acrodef{NR}{new radio}
\acrodef{LMS}{land mobile satellite}
\acrodef{IM}{intermodulation}
\acrodef{OBO}{output back-off}
\acrodef{FDD}{frequency division duplexing}
\acrodef{TDD}{time division duplexing}
\acrodef{BLER}{block error rate}
\acrodef{rv}{random variable}
\acrodef{COB}{center of beam}
\acrodef{ISL}{inter-satellite link}
\acrodef{AP}{access point}
\acrodef{SVD}{single value decomposition}
\acrodef{CW}{continuous wave}
\acrodef{AOCS}{attitude and orbit control system}
\acrodef{LSTM}{long short-term memory}
\acrodef{CPU}{central processing unit}
\acrodef{GPU}{graphics processing unit}
\acrodef{TPU}{tensor processing unit}
\acrodef{MMF}{max-min fairness}
\acrodef{MHA}{multi-head attention}
\acrodef{FFN}{feed-forward network}
\acrodef{MSE}{mean square error}
\acrodef{CDF}{cumulative distribution function}
\acrodef{EPA}{equal power allocation}
\acrodef{FPA}{fractional power allocation}
\acrodef{TNN}{transformer neural network}
\acrodef{RT}{real-time}
\acrodef{DCC}{dynamic cooperation clustering}
\acrodef{ReLU}{rectified linear unit}
\acrodef{ELU}{exponential linear unit}
\acrodef{FLOP}{Floating-point operation}

\begin{document}

\title{Linear Attention  for Joint Power Optimization and User-Centric Clustering  in  Cell-Free Networks}

\author{
\IEEEauthorblockN{Irched Chafaa,   Giacomo Bacci, \emph{Senior Member, IEEE}, Luca Sanguinetti, \emph{Fellow, IEEE}\vspace{-0.7cm}
\thanks{ 
 \newline \indent The authors are with the Dipartimento di Ingegneria dell'Informazione, University of Pisa, Via Caruso 16, 56122, Pisa, Italy (e-mail: irched.chafaa@ing.unipi.it, \{giacomo.bacci, luca.sanguinetti\}@unipi.it). This work was supported by the Smart Networks and Services Joint Undertaking (SNS JU) under the European Union’s Horizon Europe research and innovation program under Grant Agreement No 101192369 (6G-MIRAI), and by the project FoReLab (Departments of Excellence), funded by the Italian Ministry of Education and Research (MUR). L. Sanguinetti was also supported by the Project GARDEN funded by EU in NextGenerationEU Plan through Italian
“Bando Prin 2022-D.D.1409 del 14-09-2022”.
}
}}
% make the title area
\maketitle
    
%========================================================================
%                               ABSTRACT                                             
%======================================================================== 

\begin{abstract}
Optimal \ac{AP} clustering and power allocation are essential in user-centric cell-free massive \acs{MIMO} systems. However, existing solutions, whether based on deep learning or traditional optimization, lack flexibility to adapt to dynamic network configurations, neglect pilot contamination effects and incur significant computational complexity. In this paper, we propose a lightweight Transformer model that overcomes these limitations by jointly predicting \ac{AP} clusters and powers solely from the knowledge of spatial coordinates of user devices and \acp{AP}. Our model is architecture-agnostic to users load, handles both clustering and power allocation without channel estimation overhead, and eliminates pilot contamination by assigning users to \acp{AP} within a pilot reuse constraint. We also incorporate a customized linear attention mechanism to capture user-\ac{AP} interactions efficiently and enable linear scalability with respect to the number of users. Numerical results confirm the model's effectiveness in maximizing the minimum spectral efficiency and providing near-optimal performance while ensuring adaptability and scalability in dynamic scenarios. 
\end{abstract}

\vspace{0.3cm}


\acresetall
% %========================================================================
%                               INTRODUCTION                     % -*-*-*-*-
 \section{Introduction}\label{sec:intro}
\Ac{CF} \ac{mMIMO} has emerged as a promising architecture to overcome the limitations of conventional cellular networks \cite{ngo2017cell, ngo2015cell, nayebi2015cell, bjornson2019making}. Unlike traditional cellular systems, where each \ac{UE} is associated with a single base station, \ac{CF} \ac{mMIMO} deploys a large number of distributed \acp{AP} that coherently and jointly serve all \acp{UE} within the coverage area. This high level of cooperation offers uniform service quality, mitigates inter-cell interference, and enhances \ac{SE}, particularly in dense or heterogeneous environments. However, realizing these benefits in practice remains challenging. If every \ac{AP} serves every \ac{UE}, the resulting system demands substantial fronthaul signaling, creates unnecessary interference from weak \ac{AP}-\ac{UE} links, and leads to inefficient use of power and pilot resources \cite{ngo2024ultradense}.

To address these challenges, the \ac{UC} paradigm has been introduced in \ac{CF} \ac{mMIMO} \cite{demir2021foundations, buzzi2017cell, buzzi2025user, bjornson2019new}. In this approach, each \ac{UE} is served by a subset of the most relevant \acp{AP}, selected based on spatial proximity or channel quality. By restricting cooperation to the most beneficial \acp{AP}, the \ac{UC} strategy reduces interference and enhances scalability by lowering the computational and signaling burden. Nevertheless, the performance of \ac{UC} \ac{CF} \ac{mMIMO} depends critically on the proper selection of \ac{AP} clusters and transmission powers, which must adapt dynamically to variations in channel conditions, \ac{UE} mobility, and network configurations. This makes clustering and power control crucial challenges in realizing the full potential of \ac{UC} \ac{CF} \ac{mMIMO}.

 \subsection{Related work} 
  Given the importance of clustering and power optimization in \ac{UC} \ac{CF} \ac{mMIMO}, a wide range of solutions have been proposed in the literature \cite{farooq2020accelerated,demir2021foundations,miretti2022closed,shi2025joint,bjornson2013optimal,ammar2021user,di2024deep,liu2016joint,ammar2021downlink,liu2024joint}. These approaches differ in whether they treat clustering and power allocation separately or jointly, and in the methodologies employed, ranging from classical optimization to modern machine learning. 
  %In what follows, we provide a brief overview of these contributions and motivate the need for more flexible and scalable solutions.
 
Classical optimization-based methods have long been applied to power allocation, often through iterative algorithms \cite{farooq2020accelerated,demir2021foundations} or closed-form solutions for specific objectives such as the \ac{MMF} problem \cite{miretti2022closed}. While these approaches can achieve near-optimal performance, they typically require multiple iterations to converge, resulting in high computational complexity and poor scalability in dynamic networks.

In parallel, clustering strategies have been studied extensively. Heuristic approaches \cite{shi2025joint,bjornson2013optimal} provide simple rules for grouping users and \acp{AP}, while optimization-based formulations \cite{ammar2021user} aim for more principled solutions. More recently, deep learning models \cite{di2024deep} have been introduced to automate clustering decisions. Despite their advantages, these methods often lack adaptability: heuristic rules may oversimplify, optimization approaches remain computationally demanding, and deep learning solutions require retraining whenever the number of \acp{UE} or \acp{AP} changes.

Joint clustering and power allocation has also been investigated \cite{liu2016joint,ammar2021downlink,liu2024joint}, with the goal of simultaneously determining the serving \ac{AP} subset and transmission powers. These methods generally outperform separate treatments, but they still suffer from scalability issues and often neglect critical aspects such as pilot contamination. Overall, whether treated separately or jointly, existing solutions remain limited in their ability to adapt efficiently to dynamic network conditions.  

\subsection{Transformer-based power allocation}

Recently, Transformers \cite{vaswani2017attention} have been explored to address flexibility in dynamic wireless networks. In \cite{kocharlakota2024pilot}, the authors propose a Transformer-based  model only for \ac{DL} power allocation to handle varying numbers of \acp{UE} via unsupervised learning. However, the proposed method requires post-processing and padding, which introduces excessive padding when \ac{UE} loads vary widely, potentially diluting meaningful information, and increases computational load without contributing useful information \cite{dwarampudi2019effects,alrasheedi2023padding}, thereby slowing training and affecting convergence. In \cite{chafaa2025transformer}, we addressed the flexibility issue in \ac{CF} \ac{mMIMO} by leveraging a Transformer-based model capable of handling varying user densities without padding or retraining. Unlike \cite{kocharlakota2024pilot}, which employs large-scale fading coefficients as input, our model uses only spatial information (\ac{AP}/\ac{UE} coordinates) to jointly predict \ac{UL} and \ac{DL} powers. This design enables efficient adaptation to changes in the number of \acp{UE} and \acp{AP}, while maintaining near-optimal performance for the \ac{MMF} problem. Nevertheless, the solution did not address \ac{AP} clustering or pilot contamination, and its complexity still scaled quadratically with the number of users similarly to \cite{kocharlakota2024pilot}. In summary, the key question motivating this work is: \emph{How can we overcome these limitations while preserving the advantages of Transformer-based models in dynamic \ac{CF} \ac{mMIMO}?}

\subsection{Main contributions}
 {\color{black} Building on our earlier work in \cite{chafaa2025transformer}, this paper develops a scalable geometry-driven  optimization framework for \ac{UC} \ac{CF} \ac{mMIMO} systems. The proposed model learns to approximate channel-aware max-min SE optimization policies  while significantly reducing computational and signaling overhead.

The main contributions are summarized as follows:

\begin{itemize}

\item \emph{Scalable joint clustering and power control:}
We propose a single end-to-end learning framework that jointly predicts \ac{AP} clusters and \ac{UL}/\ac{DL} power  under a max-min fairness objective. Unlike prior methods that rely on repeated channel information for resource allocation, the proposed approach produces near-optimal decisions using only spatial coordinates as input, reducing overhead and decoupling clustering and power control from channel estimation and data detection. Since \ac{UE} positions capture key propagation and interference characteristics, the model learns this mapping while naturally adapting to varying \ac{UE} loads.

\item \emph{Linear-complexity attention for large-scale \ac{CF} networks:}
The proposed model employs a customized linear attention mechanism whose complexity scales linearly with the number of \acp{UE}. This makes the framework suitable for dense \ac{UC} deployments where quadratic-attention Transformers or  optimization methods become computationally prohibitive. In addition, the \ac{ELU}-based mapping improves numerical conditioning and helps preserving weaker but fairness-relevant \ac{AP}-\ac{UE} interactions in large heterogeneous networks.

\item \emph{Adaptation to varying \ac{UE} loads without architectural redesign:}
The architecture is permutation-invariant and operates with different numbers of \acp{UE} without architecture modification or retraining, addressing a key limitation of existing deep learning-based resource allocation schemes.

\item \emph{Pilot-aware architectural constraint:}
A pilot-related clustering constraint is embedded directly within the encoder, ensuring that each \ac{AP} serves no more \acp{UE} than the available orthogonal pilots. This mitigates pilot contamination effects without requiring additional pilot assignment procedures.

\item \emph{Robustness to imperfect spatial information:}
We explicitly evaluate robustness to coordinate estimation errors by injecting Gaussian perturbations into \ac{UE} positions during training and testing. The results demonstrate stable performance under input uncertainty, enhancing practical deployment feasibility.

\end{itemize} }

\subsection{Paper outline and notation}
The remainder of the paper is organized as follows. \sectionsname~\ref{sec:model} and \ref{sec:problemFormulation} describe the wireless network model and the problem formulation. \sectionname~\ref{sec:proposed} introduces the proposed \ac{ELU}-CosFormer architecture for \ac{AP} clustering and power optimization, while \sectionname~\ref{sec:training} details the training procedure. \sectionname~\ref{sec:simus} presents an extensive performance evaluation, and \sectionname~\ref{sec:complexity} provides a computational complexity analysis. Conclusions and potential future extensions are discussed in \sectionname~\ref{sec:conclusion}.

We denote the sets of real and complex numbers by $\mathbb{R}$ and $\mathbb{C}$. Matrices and vectors are written in boldface upper- and lowercase letters, respectively. {\color{black}The superscripts $(\cdot)^\Ttran$ and $(\cdot)^\Htran$ denote the transpose and the conjugate transpose operations, respectively,} and $\odot$ indicates element-wise multiplication. Element indices are written as $a_i$ for the $i$th entry of vector $\mathbf{a}$ and $a_{i,j}$ for the $(i,j)$th entry of a matrix. The notation $\mathcal{N}_{C}(\bm{\mu},\mathbf{C})$ refers to a circularly symmetric complex Gaussian variable with mean $\bm{\mu}$ and covariance matrix $\mathbf{C}$. The norm $\lVert\cdot\rVert$ denotes the $\ell_2$ vector norm, and $\mathbb{E}[\cdot]$ the expectation operator. The notation $\mathbbm{1}(\cdot)$ refers to the indicator function. {\color{black}The $N \times N$ identity matrix and the all-zero vector with $N$ elements are denoted by $\mathbf{I}_N$ and $\mathbf{0}_N$, respectively.}

 % %========================================================================
% %            System model and problem formulation                                               
% %========================================================================
\section{Wireless Network Model}\label{sec:model}

\begin{figure}[t]
  \centering
  \includegraphics[width=\columnwidth]{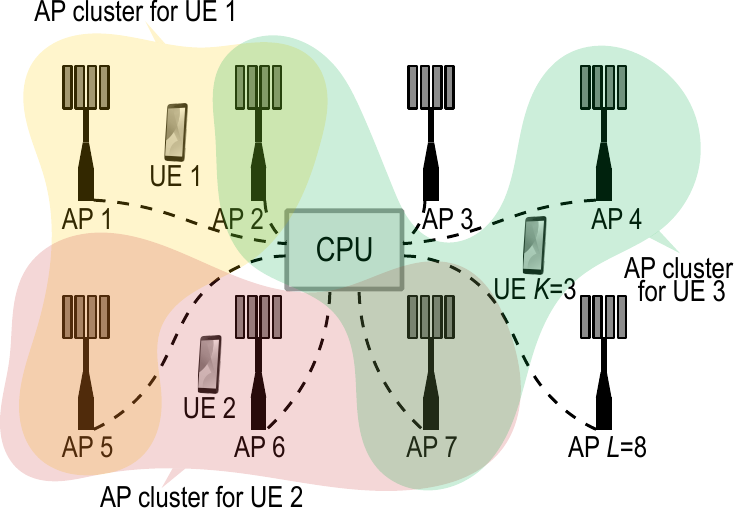}
  \caption{{\color{black}Illustration of a \acs{UC} \acs{CF} \acs{mMIMO} network, with $L=8$ \acsp{AP} and $K=3$ \acsp{UE}.} Each \acs{UE} is served by its own cluster of \acsp{AP}.}
  \label{uscf_network}
\end{figure}

We consider a \ac{UC} \ac{CF} \ac{mMIMO} network, {\color{black}like the one} shown in \figurename~\ref{uscf_network}, with $K$ single-antenna \acp{UE} and $L$ \acp{AP}, each having $N$ antennas {\color{black}(in this example, we use $K=3$, $L=8$, and $N=4$)}. Each \ac{UE} is served by a subset of \acp{AP}, {\color{black}called \emph{cluster} and indicated by the different colors in \figurename~\ref{uscf_network}}, and each \ac{AP} can serve at most $\tau_p$ \acp{UE} to prevent pilot contamination. The network operates under a standard \ac{TDD} protocol \cite{demir2021foundations}, where the $\tau_c$ symbols of each coherence block are allocated to \ac{UL} training ($\tau_p$), \ac{UL} data ($\tau_u$), and \ac{DL} data ($\tau_d$), with $\tau_c \geq \tau_p + \tau_u + \tau_d$. We adopt a narrowband channel model and assume that channels remain constant over a coherence block. The channel vector between \ac{AP} $l$ and \ac{UE} $k$ is denoted by $\mathbf{h}_{lk}$ and modeled as \cite{demir2021foundations}:
\begin{align}\label{eq:channel_model}
  \mathbf{h}_{lk} = \sqrt{\beta_{lk}} \mathbf{R}_{lk}^{1/2} \mathbf{g}_{lk},
\end{align}
where $\beta_{lk}$ is the large-scale fading, accounting for path loss and shadowing, $\mathbf{R}_{lk} \in \mathbb{C}^{N \times N}$ is the spatial correlation matrix at \ac{AP} $l$, and $\mathbf{g}_{lk} \sim \mathcal{N}_C(\mathbf{0}, \mathbf{I}_N)$ is an i.i.d. complex Gaussian vector representing the small-scale fading, where $\mathbf{I}_N$ is the $N \times N$ identity matrix. We assume that the channels $\{\mathbf{h}_{lk}; l=1,\ldots,L\}$ are independent and call $\mathbf{h}_{k} = \left[\mathbf{h}_{1k}^\Ttran, \ldots, \mathbf{h}_{Lk}^\Ttran \right]^\Ttran \in \mathbb{C}^{LN}$ the collective channel from all \acp{AP} to \ac{UE} $k$.

The \ac{CPU}, {\color{black}also shown in \figurename~\ref{uscf_network}}, computes the estimate of $\mathbf{h}_{k}$ on the basis of received pilot sequences transmitted during the training phase \cite{demir2021foundations}. The \ac{MMSE} estimate is $\widehat{\mathbf{h}}_{k} = [\widehat{\mathbf{h}}_{1k}^\Ttran, \ldots, \widehat{\mathbf{h}}_{Lk}^\Ttran ]^\Ttran$ with \cite[\sectionname~IV]{demir2021foundations}
\begin{align}
\widehat{\mathbf{h}}_{lk} = \mathbf{R}_{lk} \mathbf{Q}_{lk}^{-1} \left( \mathbf{h}_{l k} +  \frac{1}{\tau_p \rho} \mathbf{n}_{lk} \right) \sim \mathcal{N}_C \left( \mathbf{0}_N, \mathbf{\Phi}_{lk} \right),
\end{align}
where $\rho$ is the \ac{UL} pilot power of each \ac{UE}, $\mathbf{n}_{lk} \sim \mathcal{N}_C(\mathbf{0}_{N}, \sigma^2\mathbf{I}_{N})$ is the thermal noise, and $\mathbf{\Phi}_{lk} = \mathbf{R}_{lk} \mathbf{Q}_{lk}^{-1} \mathbf{R}_{lk}$, where $\mathbf{Q}_{lk} = \mathbf{R}_{l k} + \frac{\sigma^2}{\tau_p\rho} \mathbf{I}_{N}$. Hence, $\widehat{\mathbf{h}}_{k} \sim \mathcal{N}_C \left( \mathbf{0}_{LN}, \mathbf{\Phi}_{k} \right)$, with $\mathbf{\Phi}_{k} = {\mathrm{diag}}(\mathbf{\Phi}_{1k},\ldots,\mathbf{\Phi}_{Lk})$. Note that the method proposed in this paper can be applied to other channel estimation schemes.

\subsection{Uplink and downlink spectral efficiency}

An achievable \ac{UL} \ac{SE} of \ac{UE} $k$ is given by the use-and-then-forget bound \cite[\sectionname~V]{demir2021foundations}:
\begin{align}\label{eq:spectral_efficiency_uplink}
  \text{\ac{SE}}_k^\text{UL} = \frac{\tau_u}{\tau_c} \log_2\!\left(1 + \text{SINR}_k^\text{UL}\right),
\end{align}
with \ac{UE} $k$'s effective \ac{SINR} defined as
\begin{align}
\frac{p_k^\text{UL} \left| \mathbb{E}\!\left\{ \mathbf{v}_{k}^\Htran \mathbf{D}_k \mathbf{h}_{k} \right\} \right|^2}
{\sum\limits_{i=1}^K p_i^\text{UL} \mathbb{E}\!\left\{\! \left| \mathbf{v}_{k}^\Htran \mathbf{D}_k \mathbf{h}_{i} \right|^2 \!\right\}
\!-\! p_k^\text{UL} \left| \mathbb{E}\!\left\{\! \mathbf{v}_{k}^\Htran \mathbf{D}_k \mathbf{h}_{k} \!\right\} \right|^2
\!+\! \sigma^2 \mathbb{E}\!\left\{\! \| \mathbf{D}_k \mathbf{v}_k \|^2 \!\right\}}.
\end{align}
Here, $p_k^\text{UL}$ is the \ac{UL} transmit power of user $k$, $\mathbf{v}_{k}\in\mathbb{C}^N$ is the combining vector, and $\mathbf{D}_k = {\mathrm{diag}}(\mathbf{D}_{k1},\ldots,\mathbf{D}_{kL})$  is a block-diagonal matrix with $\mathbf{D}_{kl}=\mathbf{I}_N$ if \ac{AP} $l$ serves \ac{UE} $k$, and $\mathrm{diag}(\mathbf{0}_N)$ otherwise. The expectation is taken with respect to all sources of randomness. Although the bound in \eqref{eq:spectral_efficiency_uplink} is valid for any combiner $\mathbf{v}_{k}$, we consider the \ac{MMSE}, given by \cite[\sectionname~V]{demir2021foundations}: 
\begin{align}
\mathbf{v}_{k} = \left( \sum_{k=1}^{K} p_k^\text{UL} \widehat{\mathbf{h}}_{k} \widehat{\mathbf{h}}_{k}^\Htran + \mathbf{Z} \right)^{-1} \widehat{\mathbf{h}}_{k}, 
\end{align}
where
\begin{align}\mathbf{Z} =  \sum_{k=1}^{K} p_k^\text{UL}\left[\mathrm{diag}(\mathbf{R}_{1k},\ldots,\mathbf{R}_{Lk}) - \mathbf{\Phi}_{k}\right] + \sigma^2 \mathbf{I}_{LN}.
\end{align}

Similarly, the \ac{DL} \ac{SE} of user $k$ is \cite[\sectionname~VI]{demir2021foundations}
\begin{align}\label{eq:spectral_efficiency_downlink}
  \text{\ac{SE}}_k^\text{DL} = \frac{\tau_d}{\tau_c}\log_2\!\left(1 + \text{SINR}_k^\text{DL}\right),
\end{align}
where \ac{UE} $k$'s effective \ac{SINR} is
\begin{align}
\frac{p_k^\text{DL} \left| \mathbb{E}\!\left\{ \mathbf{h}_{k}^\Htran \mathbf{D}_k \mathbf{w}_{k} \right\} \right|^2}{\sum\limits_{i=1}^K p_i^\text{DL} \mathbb{E}\!\left\{ \left| \mathbf{h}_{k}^\Htran \mathbf{D}_i \mathbf{w}_{i} \right|^2 \right\} - p_k^\text{DL} \left| \mathbb{E}\!\left\{ \mathbf{h}_{k}^\Htran \mathbf{D}_k \mathbf{w}_{k} \right\} \right|^2 + \sigma^2}.
\end{align}
{\color{black}Here, $p_{k}^\text{DL}$ is the total \ac{DL} power allocated to serve \ac{UE} $k$ such that $p_{k}^\text{DL}=\sum_{l=1}^{L}{p_{k,l}^\text{DL}}$, with $p_{k,l}^\text{DL} \in \mathclose{[}0,\overline{P}_l^\text{DL}\mathclose{]}$ being \ac{AP} $l$'s transmit power allocated for user $k$; $\overline{P}_l^\text{DL}$ is the maximum power per \ac{AP} (when user $k$ is not served by \ac{AP} $l$, $p_{k,l}^\text{DL}=0$)}; and $\mathbf{w}_{k}\in \mathbb{C}^{LN}$ is the associated unit-norm precoding vector. The \ac{MMSE} precoder is used \cite{demir2021foundations}, which is given by $\mathbf{w}_{k} = \mathbf{v}_{k} / \|\mathbf{v}_{k}\|$. {\color{black} The  channel estimation and precoding/combining models adopted above are used to generate channel-aware training data as explained later in \sectionname~\ref{training:dataset}. The proposed learning framework itself does not require instantaneous  channel information at the \ac{CPU} during inference, as it predicts clustering and power decisions directly from \ac{UE}/ \ac{AP} spatial information, although channel variations are implicitly embedded during training (see \sectionname~\ref{training:supervision} for further details).}

\section{Problem Formulation}\label{sec:problemFormulation}

To ensure fairness among \acp{UE}, we adopt the \ac{MMF} problem, widely used in \ac{UC} \ac{CF} \ac{mMIMO} systems \cite{demir2021foundations, miretti2022closed, chakraborty2019centralized }. In the \ac{UL}, the problem takes the form \cite[\sectionname~VII]{demir2021foundations}:

\begin{align}
\begin{aligned}
  \max_{\{p_k^\text{UL}\geq 0\}} \min_k &\ \textrm{\ac{SE}}_k^\text{UL} \\
  \text{subject to} &\quad  p_k^\text{UL} \leq \overline{P}_k^\text{UL} \ \forall k
\end{aligned}
\label{optul}
\end{align}
where $\overline{P}_k^\text{UL}$ is the maximum \ac{UL} power for \ac{UE} $k$. {\color{black}Similarly, in the \ac{DL} we have that:
\begin{align}
\begin{aligned}
  \max_{\{p_{k,l}^\text{DL}\geq 0\}} \min_k &\ \text{\ac{SE}}_k^\text{DL}\\
  \text{subject to} &\quad \sum_{k=1}^{K} p_{k,l}^\text{DL} \leq \overline{P}_l^\text{DL} \ \forall l.
\end{aligned}
\label{optdl}
\end{align}}

{\color{black}The optimization problems in \eqref{optul} and \eqref{optdl} can be solved using a closed-form solution \cite{miretti2022closed}, iterative solvers (e.g., \cite{farooq2020accelerated,demir2021foundations}), or  deep learning models (e.g., \cite{kim2023survey,mao2018deep,kocharlakota2024pilot}) following these common steps:
\begin{enumerate}
\item Estimate the large-scale fading coefficients;
\item Assign serving \ac{AP} clusters to the \acp{UE} using a clustering algorithm (e.g., the \ac{DCC} scheme in \cite{bjornson2019new});
\item Estimate the channel vectors $\widehat{\mathbf{h}}_{l,k}$ between \acp{AP} and \acp{UE};
\item Compute the combining and precoding vectors ($\mathbf{v}_k$ and $\mathbf{w}_k$);
\item Evaluate the \ac{SE} using \eqref{eq:spectral_efficiency_uplink} or \eqref{eq:spectral_efficiency_downlink};
\item Compute the \ac{UL} and \ac{DL} transmit powers by solving \eqref{optul} and \eqref{optdl}.
\end{enumerate}

Irrespective of the adopted methodology, the solution must be obtained in real time, i.e., fast enough to be deployed before the \acp{UE}' positions, load, and spatial distribution change and the clustering and power allocation problems need to be solved again. This stringent latency requirement limits the flexibility of the system with respect to varying \ac{UE} loads and results in impractical computational overhead in large and dynamic networks, since a new solution must be computed at every coherence block. In particular, the closed-form solution in~\cite{miretti2022closed} provides the optimal solution to \eqref{optul} and \eqref{optdl} for a given fixed clustering configuration. However, it does not address the joint clustering and power allocation problem, and its computational complexity scales cubically with the number of \acp{UE}.

To address these limitations, we propose a learning model based on CosFormer \cite{qin2022cosformer}. We show that the geographical locations of \acp{UE} and \acp{AP} provide sufficient information to serve as a proxy for jointly determining the clusters and computing the optimal power allocation. We advocate using the \acp{UE}' positions since they capture the key characteristics of propagation channels and interference patterns in the network. The proposed model learns this mapping while naturally handling varying \ac{UE} loads. Once trained, the model bypasses the repeated execution of the full optimization pipeline (i.e., the six steps listed above) and produces near-optimal decisions through a single forward pass, enabling scalable and real-time adaptation to varying \ac{UE} loads while preserving fairness. 

Furthermore, this framework enables a separation between resource allocation -- potentially performed at the \ac{CPU} -- and data detection, carried out at the \acp{AP}, thereby further improving system scalability. Finally, the proposed approach can also serve as an effective system design tool, as it relies only on coarse information (i.e., the \ac{UE} locations).}

\subsection{Overview of CosFormer}

The standard Transformer \cite{vaswani2017attention} relies on an attention mechanism that captures dependencies among tokens (\acp{UE} in our case), but its quadratic complexity limits scalability. CosFormer \cite{qin2022cosformer} overcomes this issue by introducing a linearized attention mechanism that preserves accuracy while significantly reducing computational cost. Among linear Transformer variants \cite{choromanski2020rethinking,wang2020linformer,beltagy2020longformer}, {\color{black}CosFormer is particularly well suited to our setting due to the following properties:}  
\begin{enumerate}
\item \emph{Accurate and stable attention:} While many linear variants approximate the softmax attention and thus incur approximation errors, CosFormer replaces softmax with a cosine-based reweighting, improving numerical stability and avoiding such errors.
\item \emph{Implicit spatial bias:} The cosine reweighting naturally emphasizes local relationships, which aligns with the importance of nearby \ac{AP}-\ac{UE} pairs for clustering and power allocation.
\item {\color{black}\emph{Computational efficiency:} CosFormer attains linear complexity, reducing memory use and accelerating training, an essential feature for scalability in large \ac{CF} networks.} 
\end{enumerate}

Building on this foundation, we introduce our modified architecture, \emph{\ac{ELU}-CosFormer}, which augments CosFormer \cite{qin2022cosformer} with pilot-aware constraints and modified activation functions. These enhancements enable the model to jointly predict \ac{AP} clusters and powers while eliminating pilot contamination and ensuring \ac{SE} fairness. Once trained \textcolor{black}{with data that include channel modeling and estimation}, \ac{ELU}-CosFormer produces clusters and \ac{UL}/\ac{DL} power levels directly from the spatial positions of the \acp{UE} and \acp{AP}, allowing seamless adaptation to varying user loads.
% % % %========================================================================
% % %            Proposed learning model                                            
% % %===================================================================
\section{Proposed Learning Model for Clustering and Power Optimization}\label{sec:proposed}

In this section, we introduce the proposed learning model, \emph{\ac{ELU}-CosFormer}, which performs joint \ac{AP} clustering and \ac{UL}/\ac{DL} power allocation. We first describe the overall model architecture, and subsequently detail the linearized attention mechanism based on the modified CosFormer formulation.

\subsection{Model architecture}

\begin{figure}[t]
  \centering
 \includegraphics[width=0.9\columnwidth]{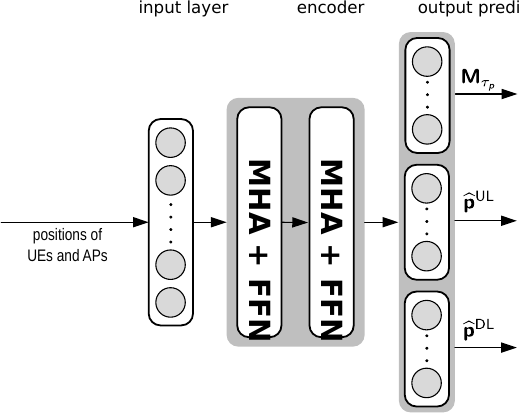}
  \caption{Architecture diagram of the proposed model to predict jointly \acs{AP} clusters, \acs{UL} and \acs{DL} powers leveraging spatial information at the input.} 
  \label{fig:arch}
  \end{figure}

The proposed model is a CosFormer-based neural network composed of three main components: a dynamic input layer, a stack of encoder layers, and parallel output heads for clustering and power prediction (see \figurename~\ref{fig:arch}). These three components are described next in detail.

\begin{figure*}[t]
  \centering
 \includegraphics[width=\textwidth]{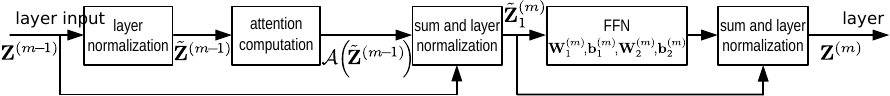}
  \caption{{\color{black}Illustration diagram of operations in one encoder layer.}} 
  \label{fig:enc}
\end{figure*}

\begin{enumerate}
\item \emph{Input layer:} The input tensor is $\mathbf{X} \in \mathbb{R}^{K \times (2L+2)}$, where each \ac{UE} is represented by $2L+2$ spatial features (coordinates of all $L$ \acp{AP} and the \ac{UE} itself). This tensor is projected into an embedding space of dimension $d_{\text{mod}}$  via a linear layer:
\begin{align}
\mathbf{Z}_{\text{in}} = \mathbf{X} \mathbf{W}_{\text{in}}^\top + \mathbf{b}_{\text{in}},
\end{align}
with trainable parameters $\mathbf{W}_{\text{in}}$ and $\mathbf{b}_{\text{in}}$. {\color{black}The input is a sequence of $K$ tokens, where each token corresponds to one \ac{UE}. Since the same projection is applied independently to every token, the operation does not assume a fixed number of \acp{UE}. Consequently, changes in $K$ only modify the sequence length, while the model parameters remain unchanged. The subsequent encoder processes this sequence using shared weights among \acp{UE} and preserves the sequence token-structure, enabling the learning model to  handle varying user loads seamlessly without requiring architectural modifications.}  
\item \emph{Encoder layers:} The embedded tensor $\mathbf{Z}^{(0)} = \mathbf{Z}_{\text{in}} \in \mathbb{R}^{K \times d_{\text{mod}}}$ is processed through $M$ encoder layers. Each layer consists of a modified CosFormer multi-head attention (MHA), denoted by $\mathcal{A}(\cdot)$, followed by a feed-forward network (FFN). Moreover, the encoder layers apply layer normalization  and residual connections to stabilize training and preserve information across layers \cite{qin2022cosformer}. Specifically, the $m$-th encoder layer applies the following operations (see \figurename~\ref{fig:enc}): 
\begin{align}
\label{eq:Z1m}
\mathbf{Z}_1^{(m)} &= \mathbf{Z}^{(m-1)} + \mathcal{A}\!\left(\tilde{\mathbf{Z}}^{(m-1)}\right),\\
\label{eq:Hm}
\mathbf{H}^{(m)} &= \mathrm{ReLU}\!\left(\tilde{\mathbf{Z}}_1^{(m)} \mathbf{W}_1^{(m)\top} + \mathbf{b}_1^{(m)}\right),\\
\label{eq:Zm}
\mathbf{Z}^{(m)} &= \tilde{\mathbf{Z}}_1^{(m)} + \mathbf{H}^{(m)} \mathbf{W}_2^{(m)\top} + \mathbf{b}_2^{(m)},
\end{align}
where $\tilde{\mathbf{Z}}$ denotes a layer-normalized version of a tensor $\mathbf{Z}$; $\mathbf{W}_1^{(m)}$, $\mathbf{b}_1^{(m)}$, $\mathbf{W}_2^{(m)} $, and $\mathbf{b}_2^{(m)}$ are  trainable parameters; $\mathrm{ReLU}$ refers to an activation function \cite{rasamoelina2020review} introducing a non-linearity.

All $M$ layers share the same structure, and the output of each encoder layer preserves the dimension $\mathbb{R}^{K \times d_{\text{mod}}}$, serving as input to the next layer. After $M$ layers, the final tensor $\mathbf{Z}^{(M)} = \mathbf{Z}_{\text{out}}$ represents the encoder output that will be processed next by the output heads. The attention mechanism in \eqref{eq:Z1m}-\eqref{eq:Zm} is detailed in \sectionname~\ref{subsec:attention}. 
\item \emph{Output heads:} Three parallel heads operate on the final encoder output $\mathbf{Z}_{\text{out}} \in \mathbb{R}^{K \times d_{\text{mod}}}$:

\begin{itemize}
    \item \emph{\Ac{UC} clustering mask}: A fully-connected layer with sigmoid activation \cite{rasamoelina2020review} produces an initial mask tensor $\mathbf{M} \in \mathbb{R}^{K \times L}$, where each entry represents the association strength between a \ac{UE} and an \ac{AP}:
    \begin{align}
    \mathbf{M} = \mathrm{sigmoid}\left(\mathbf{Z}_{\text{out}} \mathbf{W}_{\text{mask}} + \mathbf{b}_{\text{mask}}\right),
    \end{align}
    with trainable parameters $\mathbf{W}_{\text{mask}}$ and $\mathbf{b}_{\text{mask}}$.

    To mitigate pilot contamination, a two-step procedure is applied. First, for each \ac{AP}, only the top $\tau_p$ \acp{UE} with the highest mask scores are retained via $\mathrm{Top}_{\tau_p}(\mathbf{M})$. Then, a thresholding operation assigns the \ac{AP} to serve the \ac{UE} if the retained score exceeds a threshold $\xi$, yielding the final binary clustering mask:
    \begin{align}
    \mathbf{M}_{\tau_p} = \mathbbm{1}\left( \mathrm{Top}_{\tau_p}(\mathbf{M}) \odot \mathbf{M} > \xi \right).
    \end{align}
    The threshold $\xi$ is a hyperparameter tuned during training to balance connectivity and interference. Proper selection of $\xi$ ensures strong \ac{UE}-\ac{AP} associations while maintaining high \ac{SE}.

    \item \emph{Uplink power prediction}: A fully-connected layer with sigmoid activation predicts normalized \ac{UL} powers:
    \begin{align}
    \tilde{p}_k^{\text{UL}} = (\mathrm{sigmoid}\left(\mathbf{Z}_{\text{out}} \mathbf{W}_{\text{UL}} + \mathbf{b}_{\text{UL}}\right))_k,
    \end{align}
    where $\mathbf{W}_{\text{UL}}$ and $\mathbf{b}_{\text{UL}}$ are trainable parameters. The denormalized \ac{UL} powers are obtained as
\begin{align}
\hat{p}_k^{\text{UL}} = \Delta_{\text{UL}} \tilde{p}_k^{\text{UL}} + \underline{P}_{\text{UL}}, \label{eq:hat_UL}
\end{align}
where $\Delta_{\text{UL}} = \overline{P}_{\text{UL}} - \underline{P}_{\text{UL}}$ denotes the \ac{UL} power range. Here, $\overline{P}_{\text{UL}}$ and $\underline{P}_{\text{UL}}$ are the maximum and minimum allowable UL transmit powers.
    
{\color{black} \item \emph{Downlink power prediction}: A fully-connected layer with sigmoid activation outputs normalized per-\ac{AP}  \ac{DL} power weights:  
\begin{align}
\tilde{p}_{k,l}^{\mathrm{DL}} =
(\mathrm{sigmoid}\!\left(\mathbf{Z}_{\mathrm{out}}\mathbf{W}_{\mathrm{DL}}+\mathbf{b}_{\mathrm{DL}}\right))_{k,l},
\end{align}
where $\mathbf{W}_{\mathrm{DL}}$ and $\mathbf{b}_{\mathrm{DL}}$ are trainable parameters, and $\tilde{p}_{k,l}^{\mathrm{DL}}\in[0,1]$ represents the relative power weight assigned by \ac{AP} $l$ to \ac{UE} $k$.}
\end{itemize}

 {\color{black} To enforce the per-\ac{AP}  power constraint in \eqref{optdl}, the predicted weights are scaled by the \ac{AP} power budget:
\begin{align}
\hat{p}_{k,l}^{\mathrm{DL}} =
\frac{\tilde{p}_{k,l}^{\mathrm{DL}}}{\sum_{i=1}^{K}\tilde{p}_{i,l}^{\mathrm{DL}}}
\,\overline{P}_l^{\mathrm{DL}}.
\end{align} 
Since the available \ac{DL} power labels are defined per \ac{UE}, the total predicted \ac{DL} power allocated to \ac{UE} $k$ is computed as $\hat{p}_k^{\mathrm{DL}} = \sum_{l=1}^{L} \hat{p}_{k,l}^{\mathrm{DL}} $. These values are then normalized using the min-max scaling criterion introduced in \sectionname~\ref{training:dataset} (Step 3) to obtain the predicted labels used for training. This ensures that the learning model respects the per-\ac{AP} power constraints while remaining consistent with the available optimal labels used in the loss function defined in \sectionname~\ref{training:supervision}.}

\end{enumerate}

\subsection{Attention mechanism}\label{subsec:attention}

We now detail the attention mechanism used in \eqref{eq:Z1m}-\eqref{eq:Zm}. The queries ($\mathbf{Q}$), keys ($\mathbf{K}$), and values ($\mathbf{V}$) \cite{vaswani2017attention} are simultaneously computed from the layer-normalized input $\tilde{\mathbf{Z}}_{\text{in}}$ using a single linear transformation:
\begin{align}
 \mathbf{\Omega} = \tilde{\mathbf{Z}}_{\text{in}} \mathbf{W}_{\mathbf{\Omega}} + \mathbf{b}_{\mathbf{\Omega}},
\end{align}
where $\mathbf{W}_{\mathbf{\Omega}} \in \mathbb{R}^{d_{\text{mod}} \times 3d_{\text{mod}}}$ and $\mathbf{b}_{\mathbf{\Omega}} \in \mathbb{R}^{3d_{\text{mod}}}$ are trainable parameters. The resulting tensor $\mathbf{\Omega} \in \mathbb{R}^{B \times K \times 3d_{\text{mod}}}$ is split along the last dimension into three parts, yielding $\mathbf{Q}, \mathbf{K}, \mathbf{V} \in \mathbb{R}^{B \times K \times d_{\text{mod}}}$. These are then reshaped into $N_h$ attention heads of dimension $d_{\text{head}} = d_{\text{mod}} / N_h$, giving
\begin{align}
    \mathbf{Q}, \mathbf{K}, \mathbf{V} \in \mathbb{R}^{B \times N_h \times K \times d_{\text{head}}}.
\end{align}
A kernel feature mapping $\phi(\cdot)$ is applied to the queries and keys:
\begin{align}
\mathbf{Q}' &= \phi(\mathbf{Q}), \\
\mathbf{K}' &= \phi(\mathbf{K}),
\label{phi_eq}
\end{align}
with $\phi(\cdot)$ being an activation function chosen to approximate the softmax kernel \cite{rasamoelina2020review} of the original Transformer while enabling linearized attention. 

The attention is then computed via a linearized operation:
\begin{align}
\mathbf{A} = \mathbf{Q}' \cdot \Big( \big( \mathbf{K}'^\top \cdot \mathbf{V} \big) \cdot \mathbf{W} \Big),
\label{eq:lin}
\end{align}
where $\mathbf{W} \in \mathbb{R}^{K \times K}$ is a fixed cosine reweighting matrix with entries $w_{ij} = \cos\!\left( \pi(i - j)/({2K}) \right)$. This modulation biases attention toward nearby tokens, mimicking the locality bias of softmax without computing full pairwise interactions, thereby avoiding the quadratic complexity. In our context, this allows the model to distinguish between nearby and distant \acp{UE}/\acp{AP} without introducing additional trainable parameters.

Finally, $\mathbf{A}$ is projected back into the model dimension, yielding the final attention weights used in \eqref{eq:Z1m}:
\begin{align}
\label{eq:Zin}
\mathcal{A}\!\left(\tilde{\mathbf{Z}}^{(m-1)}\right) = \mathbf{A} \mathbf{W}_{\text{out}}^\top + \mathbf{b}_{\text{out}},
\end{align}
with $\mathbf{W}_{\text{out}} \in \mathbb{R}^{d_{\text{mod}} \times d_{\text{mod}}}$ and $\mathbf{b}_{\text{out}} \in \mathbb{R}^{d_{\text{mod}}}$ being trainable parameters.

\emph{Our choice of  activation function}. In the original CosFormer \cite{qin2022cosformer}, the kernel feature mapping function $\phi(\cdot)$ is implemented using  \ac{ReLU}. In this work, we propose a simple yet efficient modification: replacing \ac{ReLU} with the \ac{ELU} \cite{rasamoelina2020review}. This choice is motivated by both physical insights and empirical evidence. Formally, the two activations are defined as \cite{rasamoelina2020review}:
\begin{align}
\label{activation}
\begin{aligned}
\mathrm{ReLU}(x) &= \max(0, x),\\
\mathrm{ELU}(x) &= \begin{cases}
x & \textrm{if $x > 0$},\\
\alpha (e^x - 1) & \textrm{if $x \leq 0$},
\end{cases}
\end{aligned}
\end{align}
where $\alpha$ is a hyperparameter, typically set to $1$.

{\color{black}\ac{ELU} provides smoother gradients and nonzero outputs for negative inputs, preserving richer feature interactions and improving gradient flow and numerical stability during training. This is particularly important in our setting, where spatial coordinates require the attention mechanism to capture subtle \ac{UE}-\ac{AP} variations. In contrast, \ac{ReLU} zeroes out all negative inputs, potentially discarding useful spatial information and leading to sparse gradients that hinder convergence. By deliberately allowing negative values, \ac{ELU} maintains continuity across the input domain and prevents inactive attention heads, enabling the attention mechanism to exploit weak but meaningful \ac{AP}-\ac{UE} associations that would otherwise be suppressed. Such associations are critical under \ac{MMF} criterion, where low-gain links can influence the performance of the worst-case \ac{UE}.} 

{\color{black} Moreover, the exponential tail of \ac{ELU} produces a smooth and fully differentiable activation, which interacts naturally with the cosine modulation in \eqref{eq:lin} (through $\bf{W}$). This combination yields a differentiable kernel that results in smoother and more stable attention distributions, reducing abrupt variations in the attention weights. This behavior is consistent with the inherently smooth spatial variations observed in wireless propagation, where signal strengths change gradually rather than abruptly. Consequently, the learned attention patterns better reflect realistic relationships between \acp{UE} and \acp{AP}, leading to more physically meaningful clustering boundaries and a more balanced power allocation across the network. The effectiveness of this activation function is evaluated in \sectionname~\ref{sec:simus} via an ablation experiment.}

\section{Training Setup}\label{sec:training}
The model is trained offline to predict the \ac{AP} clusters and \ac{UE} powers. Next, we explain the structure of the generated dataset and provide details about the training process.

\subsection{Dataset Construction}\label{training:dataset}

{\color{black}To train and evaluate the learning model, we construct, based on the model in \sectionname~\ref{sec:model}, a synthetic dataset that captures diverse cell-free network configurations and realistic deployment conditions.}  The dataset construction follows four main steps:

\begin{enumerate}[label=\textbf{Step \arabic*)}, leftmargin=*, align=left, labelsep=0.5em]
\item \textcolor{black}{ \emph{Data generation:} For each configuration defined by a pair $(K, L)$, we generate:}
    \begin{itemize}
        \item \textcolor{black}{random 2D positions within a bounded area for $K$ single-antenna \acp{UE}, denoted as $\{\mathbf{u}_k\}_{k=1}^K$, with $\mathbf{u}_k \in \mathbb{R}^2$, and $L$ uniformly distributed \acp{AP}, denoted as $\{\mathbf{a}_l\}_{l=1}^L$, with $\mathbf{a}_l \in \mathbb{R}^2$;}
        \item  \textcolor{black}{large-scale fading coefficients $\beta_{lk}$, correlation matrices $\mathbf{R}_{lk}$, \ac{MMSE} estimated channels $\mathbf{\widehat{h}}_{lk}$, \ac{MMSE} precoders $\mathbf{w}_{k}$ and combiners $\mathbf{v}_{k}$  computed as detailed in \sectionname~\ref{sec:model}, which are required later by the closed-form solution to solve the max-min \ac{SE} optimization problems in \eqref{optul} and \eqref{optdl} for each set of clusters.}
    \end{itemize}
\item \emph{Noise injection:} To simulate localization errors and improve robustness, Gaussian noise is added to \ac{UE} positions:
    \begin{align}\label{eq:trainingNoise}
        \mathbf{u}_k' &= \mathbf{u}_k + \delta_k, \quad \delta_k \sim \mathcal{N}(0, \sigma_e^2),
    \end{align}
    where $\sigma_e = 1\meter$, following \cite{xue2021analysis}. This perturbation accounts for position estimation errors.
\item \emph{Normalization:} All features are normalized to the range $[0,1]$ using min-max scaling. For a feature vector $\boldsymbol{\xi}$, we use
    \begin{align}
    \label{norm}
        \boldsymbol{\tilde{\xi}} = \frac{\boldsymbol{\xi} - \min(\boldsymbol{\xi})}{\max(\boldsymbol{\xi}) - \min(\boldsymbol{\xi}) + \varepsilon},
    \end{align}
    where $\varepsilon$ is a small constant to avoid division by zero. Coordinates are normalized separately for 
$x$ and $y$, preserving geometric relationships, while \ac{UL} and \ac{DL} powers are normalized globally to ensure consistent scaling.

\begin{algorithm}[t]
\caption{Training with Dynamic Supervision}
\label{alg:training}
\begin{algorithmic}[1]
\Require Training dataset $\{\mathbf{X}_i\}_{i=1}^S$, batch size $S$, trade-off parameter $\lambda$
\For{each epoch}
\For{each batch}
\State Compute attention scores from the input $\mathbf{X}_i$ 
\State Generate \ac{AP} clusters for each \ac{UE} 
\State Compute powers by solving \eqref{optul} and \eqref{optdl} \cite{miretti2022closed}
\State Evaluate the loss function \eqref{eq:loss}
\State Update model parameters via backpropagation
\EndFor
\EndFor
\end{algorithmic}
\end{algorithm}
\item \emph{Dataset overview:} A total of $8000$ training samples are generated for each value of $K \in \{5,10\}$ and $L =16$. Each sample consists of normalized noisy \ac{UE} and \ac{AP} coordinates, with corresponding optimal powers  \emph{computed on the fly during training}. To evaluate generalization, additional samples are generated for other user counts beyond the training values of $K$, enabling assessment of the model’s ability to adapt to unseen network configurations. Overall, the dataset includes diverse configurations by varying the number of \acp{UE}, their spatial distributions, and channel realizations, ensuring robust learning and effective generalization.
\end{enumerate}

\subsection{Training approach}\label{training:supervision}
{\color{black}
Unlike conventional supervised learning \cite{nasteski2017overview}, the optimal power labels cannot be precomputed prior to training because the clusters of \acp{AP} serving each \ac{UE} are determined dynamically by the model itself. Consequently, the optimal power labels depend on the clustering decisions produced during the forward pass and are computed on-the-fly during training. This results in a supervised learning framework with dynamic label generation, also referred to as dynamic supervision \cite{pham2021meta,wang2020learning}, where the supervision signals are obtained by solving the power control problems for the clusters predicted by the model.}

The overall training procedure is summarized in \algoname~\ref{alg:training}. During the forward pass, the model evaluates attention scores between the input features, which define the \ac{AP} clusters by assigning a set of serving \acp{AP} to each \ac{UE}. Given these predicted clusters, the \ac{UL} and \ac{DL} power control problems are solved using the closed-form expressions in \cite{miretti2022closed}. Then, the predicted powers $\tilde{\mathbf{p}}$ are  compared with the normalized optimal ones $\tilde{\mathbf{p}}^{\star}$, and the loss function updates the model parameters:
\begin{align}
\!\!\!\!\mathcal{L} = \frac{1}{S} \sum_{i=1}^S \|\tilde{\mathbf{p}}^{\star}_i - \tilde{\mathbf{p}}_i\|^2 
  - \lambda \left( \min_k \textrm{\ac{SE}}_k^\text{UL} + \min_k \textrm{\ac{SE}}_k^\text{DL}\right),
  \label{eq:loss}
\end{align}
where $S$ is the batch size. The loss combines the mean-squared error between predicted and optimal powers with a clustering-aware penalty that promotes fairness across users. The hyperparameter $\lambda$ controls the trade-off between minimizing prediction error and improving the worst-case \ac{SE}. In practice, $\lambda$ is tuned empirically via cross-validation.

The model is implemented in PyTorch \cite{paszke2019pytorch} and configured with the parameters listed in \tablesname~\ref{tab:arch_params} and \ref{tab:train_params}, which are empirically tuned to balance accuracy and computational efficiency.

{\color{black}Notably, the model does not take channel coefficients as input; instead, their effect is  captured through the optimal power labels. By learning from these labels, the network maps \ac{UE}-\ac{AP} spatial configurations to clustering and power allocation decisions across different network configurations and user loads, effectively approximating the max-min \ac{SE} policy at inference time without requiring instantaneous channel information.
}

\begin{table}[t]
\centering
\caption{Model architecture parameters.}
\label{tab:arch_params}
\begin{tabular}{ll}
\hline
\textbf{Parameter} & \textbf{Value} \\
\hline
Encoder layers $M$ & $2$ \\
Attention heads $N_h$ & $4$ \\
Model dimension $d_{\text{mod}}$ & $64$ \\
Clustering threshold $\xi$ & $0.3$ \\
\hline
\end{tabular}
\end{table}

\begin{table}[t]
\centering
\caption{Model training hyperparameters.}
\label{tab:train_params}
\begin{tabular}{ll}
\hline
\textbf{Parameter} & \textbf{Value} \\
\hline
Trade-off parameter $\lambda$ & $10^{-2}$ \\
Dropout rate & $0.1$ \\
Optimizer & AdamW \cite{zhou2024towards} \\
Learning rate & $10^{-3}$ \\
Epochs & $20$ \\
Batch size & $64$ \\
\hline
\end{tabular}
\end{table}

%%%%%%%%%%%%%%%%%%%%%%%%%%%%%%%%%%%%%%%%%%%%%%%%%%%%%%%%%%%%%%
 %                               SIMULATION RESULTS
%%%%%%%%%%%%%%%%%%%%%%%%%%%%%%%%%%%%%%%%%%%%%%%%%%%%%%%%%%%%%%%%%%%%%%%%%%%%%%%%%%%

\section{Numerical results}\label{sec:simus}
In this section, we assess the performance of the proposed learning-based framework for solving the \ac{MMF} problem in the \ac{UC} \ac{CF} \ac{mMIMO} system described in \sectionname~\ref{sec:model}. The main simulation parameters are reported in Table~\ref{tab:sim_params}. All numerical results are averaged over the test set.

The proposed solution is compared against the following clustering and power allocation schemes:
\begin{itemize}
\item \emph{Optimal \acs{CF}:} A fully cooperative baseline in which all \acp{AP} serve all \acp{UE}, and transmit powers are obtained via the closed-form expression in \cite{miretti2022closed}.
\item \emph{\acs{DCC}:} A fixed-clustering benchmark based on the \ac{DCC} scheme \cite{bjornson2019new}, where each \ac{UE} is associated with its $Q=8$ strongest \acp{AP}. Power control follows the closed-form solution.
\item \emph{Optimal \acs{UC}-\acs{CF}:} A hybrid strategy in which \ac{AP}-\ac{UE} clusters are inferred by the proposed model, while transmit powers are still computed using the closed-form solution.
\item \emph{Predicted \acs{UC}-\acs{CF}:} The fully data-driven approach, where both clustering decisions and power levels are directly predicted by the proposed learning architecture.
\end{itemize}

\begin{table}[t]
\centering
\caption{\Ac{CF} \ac{mMIMO} network parameters.}
\label{tab:sim_params}
\begin{tabular}{ll}
\hline
\textbf{Parameter} & \textbf{Value} \\
\hline
Network area & $500\meter \times 500\meter$ \\
Number of \acp{AP} ($L$) & $16$ (uniformly deployed) \\
Antennas per \acs{AP} ($N$) & $4$ \\
Carrier frequency & $2\GHz$ \\
Path-loss exponent & $3.67$ \\
\acs{UE}-\acs{AP} height difference & $10\meter$ \\
Shadow fading $F_{kl}$ & $\mathcal{N}_{C}(0, \alpha^2), \; \alpha^2 = 4\dB$ \\
% Shadow fading correlation & As in \cite[\sectionname~III-D]{miretti2022closed} \\
Noise power $\sigma^2$ & $-94\dB$ \\
Noise figure $\eta$ & $7\dB$ \\
Bandwidth $B$ & $20\MHz$ \\
Max \acs{UL} transmit power $\overline{P}_\text{UL}$ &   $100\mW$ \\
Max \acs{DL} transmit power per \acs{AP} $\overline{P}_l^\text{DL}$ &  $200\mW$ \\
Coherence block length $\tau_c$ & $200$ \\
Pilot length $\tau_p$ & $10$ \\
\acs{UL} data symbols $\tau_u$ & $90$ \\
\acs{DL} data symbols $\tau_d$ & $100$ \\
\hline
\end{tabular}
\end{table}

\subsection{{\color{black}Activation function performance}}

\begin{figure}[t!]
    \centering
    \begin{subfigure}[b]{\columnwidth}
        \centering
        \includegraphics[width=\textwidth]{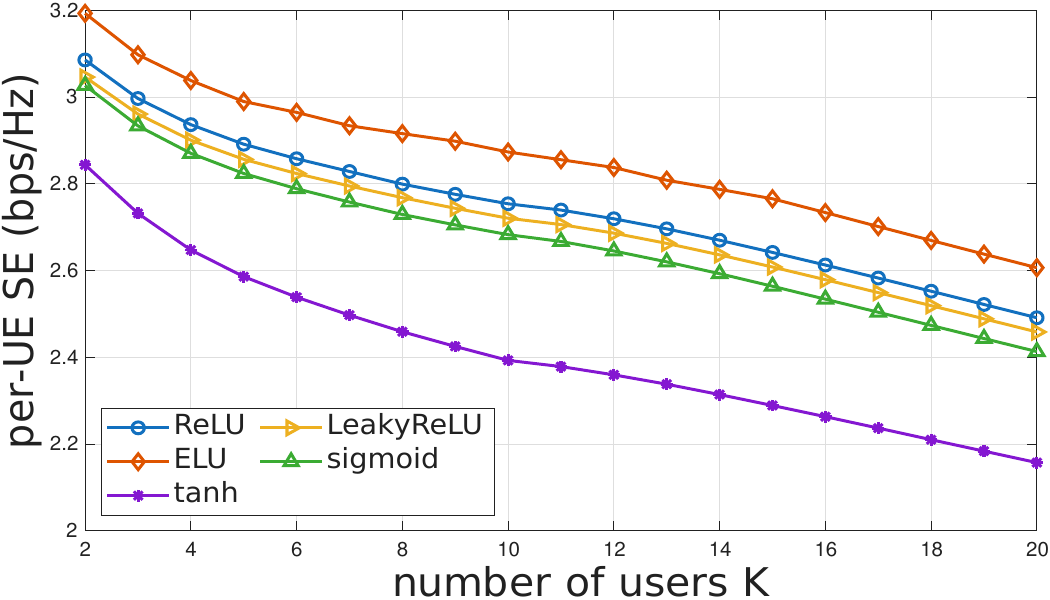} 
        \caption{\acs{UL}}
        \label{fig:ul_se_kernel}
    \end{subfigure}
    \hfill
    \begin{subfigure}[b]{\columnwidth}
        \centering
        \includegraphics[width=\textwidth]{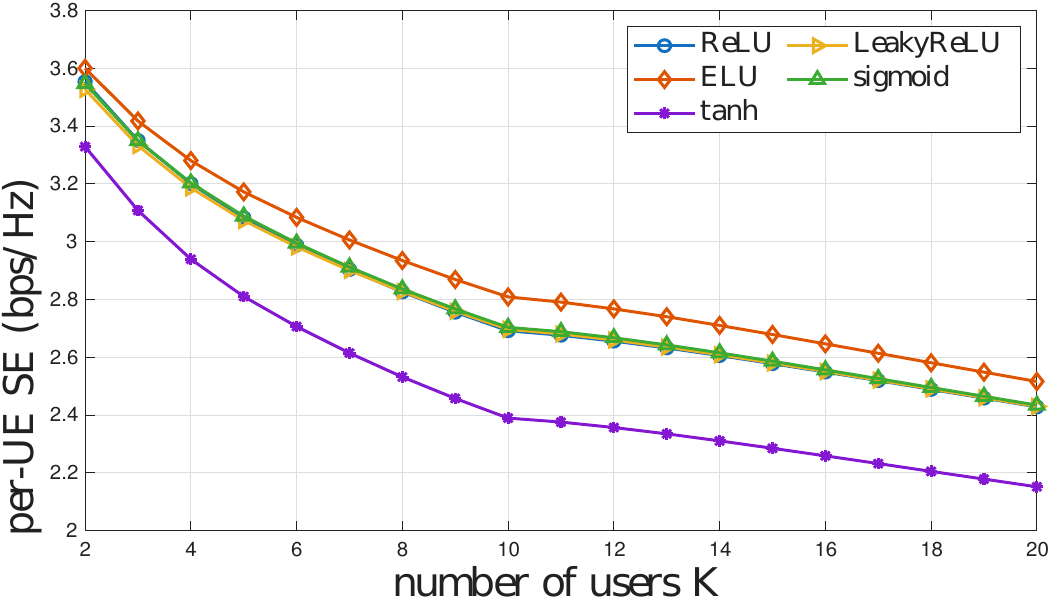}
        \caption{\acs{DL}}
        \label{fig:dl_se_kernel}
    \end{subfigure}
    \caption{Comparison of per-\acs{UE} \ac{SE} using different activation functions in CosFormer for \acs{UL} and \acs{DL}. \acs{ELU} yields the highest \acs{SE} across all \acs{UE} counts.}
    \label{fig:se_kernel}
\end{figure}

{\color{black}To validate the choice of the \ac{ELU} activation function detailed in \sectionname~\ref{subsec:attention}, we evaluate the per-\ac{UE} \ac{SE} achieved by a CosFormer using several  activation functions \cite{rasamoelina2020review} for the attention computation with the simulation parameters in \tablename~\ref{tab:sim_params}. The results reported in \figurename~\ref{fig:se_kernel} demonstrate that \ac{ELU} consistently yields higher \ac{SE} across all user counts $K$ for both \ac{UL} and \ac{DL} scenarios, confirming that \ac{ELU} provides a more effective attention computation for our problem.}

\subsection{Spectral efficiency evaluation}
\begin{figure}[htbp]
    \centering
    \begin{subfigure}[b]{\columnwidth}
        \centering
        \includegraphics[width=\textwidth]{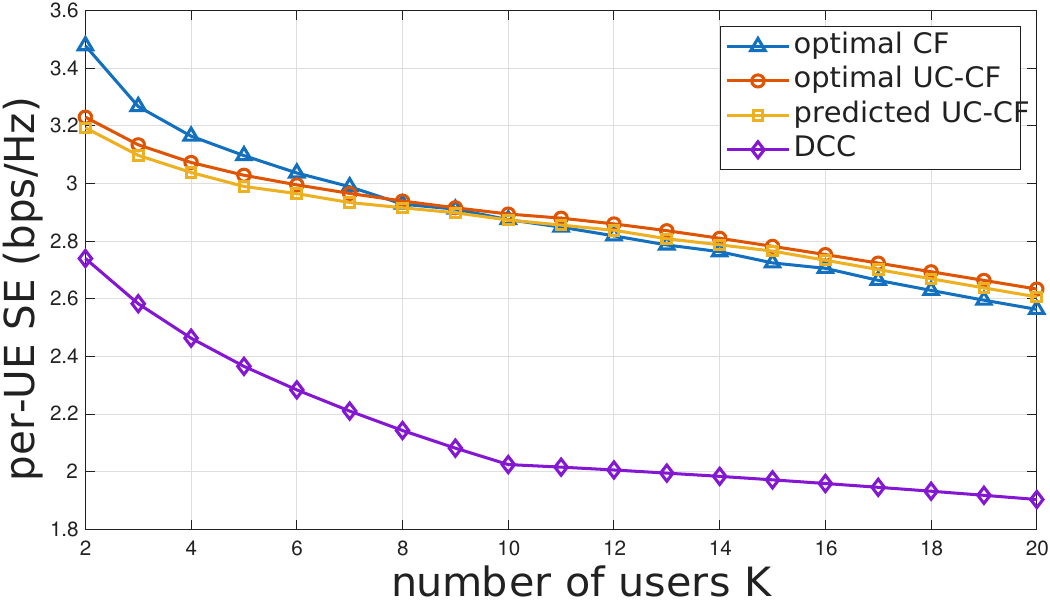}
        \caption{\acs{UL}}
        \label{fig:ul_se}
    \end{subfigure}
    \hfill
    \begin{subfigure}[b]{\columnwidth}
        \centering
        \includegraphics[width=\textwidth]{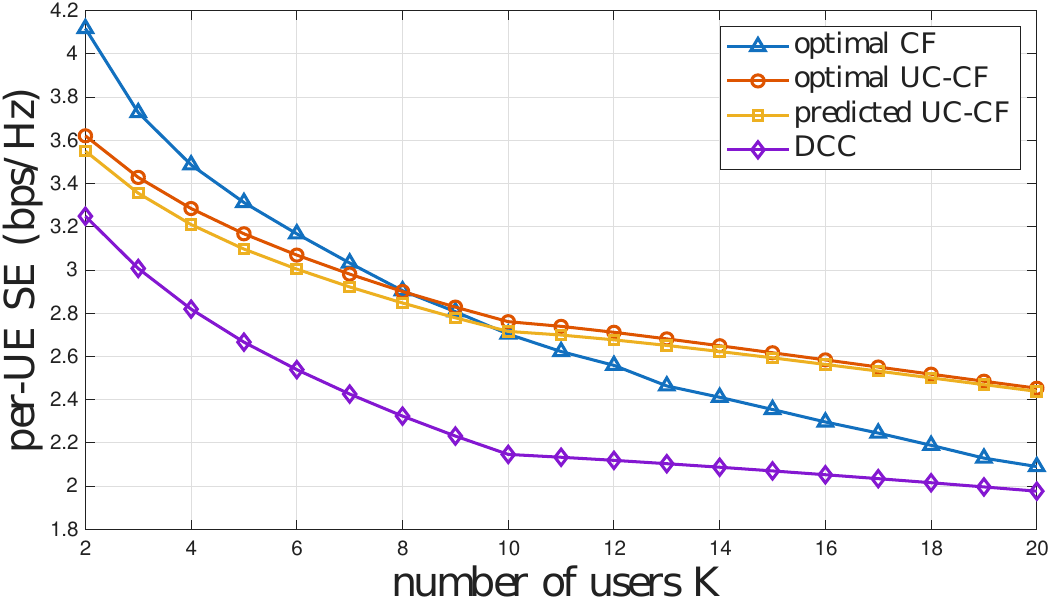}
        \caption{\acs{DL}}
        \label{fig:dl_se}
    \end{subfigure}
    \caption{Average per-\acs{UE} \acs{SE} in \acs{UL} and \acs{DL} for different $K$ values on the test set. The proposed model adapts better to increasing user load compared to benchmark policies.}
    \label{fig:se}
\end{figure}

\figurename~\ref{fig:se} reports the average per-\ac{UE} \ac{SE} as a function of the number of users $K$ for both \ac{UL} and \ac{DL} transmission. As expected, the per-\ac{UE} \ac{SE} decreases with increasing $K$ in both links, since a larger number of users must share a fixed amount of \ac{AP} resources and is subject to higher interference. Nevertheless, the rate at which performance degrades varies significantly across the considered strategies.
In the \ac{UL} (\figurename~\ref{fig:ul_se}), \emph{Optimal \acs{CF}} provides the highest \ac{SE} for small values of $K$ due to full \ac{AP} cooperation. However, its performance deteriorates more rapidly as $K$ increases. Beyond $K \approx 10$, \emph{Optimal \acs{UC}-\acs{CF}} surpasses the fully cooperative baseline, as limiting cooperation to the most relevant \acp{AP} reduces interference and enables a more effective allocation of transmit power. A similar behavior is observed in the \ac{DL} (\figurename~\ref{fig:dl_se}): while full \ac{CF} remains superior at low user densities, \emph{Optimal \acs{UC}-\acs{CF}} achieves higher \ac{SE} for larger $K$, with the benefits of clustering being even more pronounced in \ac{DL} due to improved power concentration and reduced inter-user interference. The \emph{Predicted \acs{UC}-\acs{CF}} closely tracks the optimal \ac{UC}-\ac{CF} in both \ac{UL} and \ac{DL}, confirming the generalization capability of the trained model across unseen user loads. In contrast, the classical \ac{DCC} consistently yields the lowest \ac{SE}, as its fixed clustering cannot adapt to varying interference conditions.  

Overall, these results show that while fully \ac{CF} operation is advantageous at low \ac{UE} densities, \ac{UC} clustering becomes increasingly beneficial as the network grows congested. The proposed learning model adapts to this transition, sustaining near-optimal performance with reduced computational and signaling overhead.

\subsection{\acs{AP}-\acs{UE} clustering performance}

\begin{figure}[t]
  \centering
 \includegraphics[width=\columnwidth]{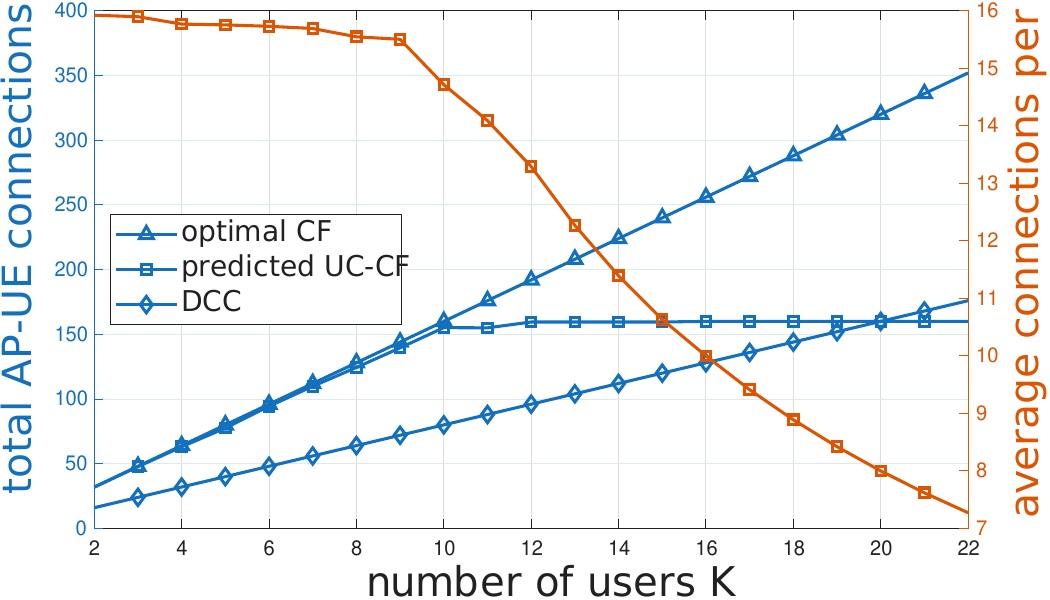}
  \caption{Total \acs{AP}-\acs{UE} connections (left axis) and average connections per \acs{UE} (right axis) as functions of different user loads. Our model in \acs{UC}-\acs{CF} strategy adapts the clusters dynamically to handle the increasing user load.}
  \label{connect}
  \end{figure}

In \figurename~\ref{connect}, we evaluate the total number of \ac{UE}/\ac{AP} connections (when \ac{UE} is served by \ac{AP}) of optimal \ac{CF}, predicted \ac{UC}-\ac{CF} and \ac{DCC} schemes for different user loads. In addition, we plot, in the secondary Y-axis, the average number of connections per \ac{UE} of the predicted \ac{UC}-\ac{CF}.

The results show that the \ac{UC}-\ac{CF} scheme exhibits fewer total connections compared to \ac{CF}, reflecting the selective association of users to a subset of \acp{AP} based on the learned clustering. This indicates an efficient usage of network resources since less connections implies less signaling and power consumption while still providing near-optimal \ac{SE} as illustrated earlier in \figurename~\ref{fig:se}. Moreover, it can be seen that the total \ac{AP}-\ac{UE} connections saturates for higher user loads as the \acp{AP} reach their $\tau_p$-served  \ac{UE} constraint unlike other policies that keep increasing, affecting their scalability. Correspondingly, the average number of \acp{AP} per \ac{UE} (red curve) gradually decreases as $K$ increases. This behavior confirms the model’s ability to dynamically reconfigure clusters and allocate \ac{AP}  efficiently under varying user densities to achieve better scalability. In contrast, the \ac{DCC}, which always connects the \ac{UE} to $Q=8$ \acp{AP}, yields the low total connections at first, but at the expense of a poor \ac{SE} performance (\figurename~\ref{fig:se}) and a linear growth  of \ac{AP}-\ac{UE} connections with $K$. 

\begin{figure}[t]
  \centering
  \includegraphics[width=0.7\columnwidth]{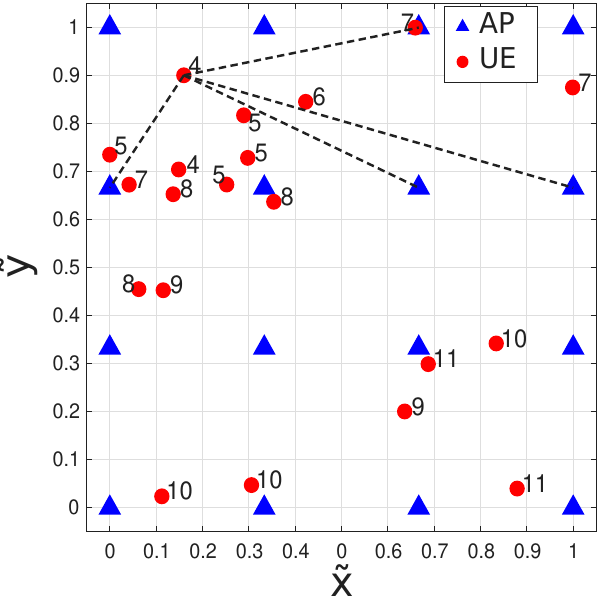}
  \caption{Network layout of \acp{AP} and \acp{UE} for $K=20$ users. User-specific clusters are predicted by the model for efficient resource allocation.}
  \label{fig:apue_layout}
\end{figure}

\figurename~\ref{fig:apue_layout} represents an example of the clustering behavior of the  model for $K=20$ and $L=16$ for a given sample of the testing set. Each \ac{UE} is annotated with the number of \acp{AP} it is connected to, revealing the model’s adaptive clustering. Dashed lines highlight the \ac{AP}-\ac{UE} links, showing selective association to only the most relevant \acp{AP}. This visualization confirms that \ac{UC}-\ac{CF} forms compact, user-specific clusters that vary across users, enabling efficient resource allocation and reduced signaling overhead.

\subsection{Power prediction accuracy}
\begin{figure}[t!]
    \centering
    \begin{subfigure}{\columnwidth}
        \centering
        \includegraphics[width=\linewidth]{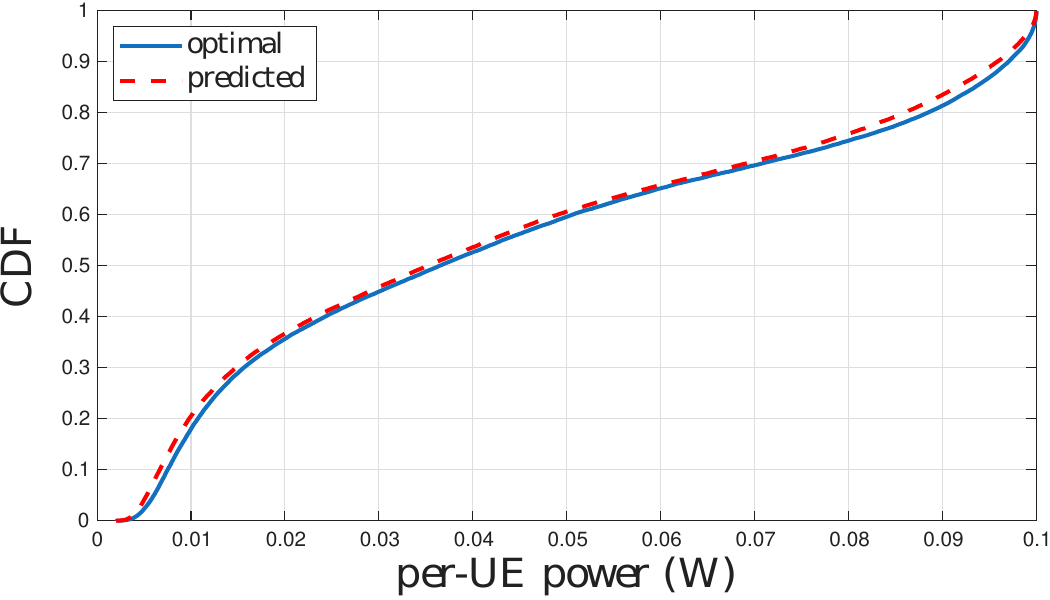}
        \caption{\acs{CDF} of \acs{UL} transmit powers.}
        \label{fig:cdf_ul_powers}
    \end{subfigure}
    \hfill
    \begin{subfigure}{\columnwidth}
        \centering
        \includegraphics[width=\linewidth]{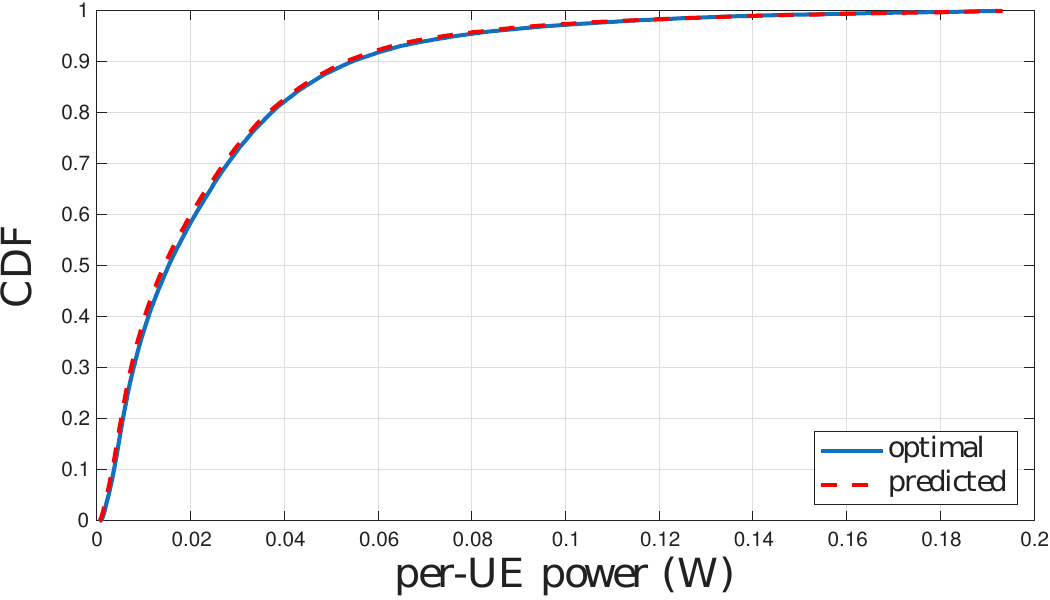}
        \caption{\acs{CDF} of \acs{DL} transmit powers.}
        \label{fig:cdf_dl_powers}
    \end{subfigure}
    \caption{\acs{CDF} comparison of optimal and predicted powers in \acs{UL} and \acs{DL}. The close overlap confirms the accuracy of the proposed learning model in reproducing the optimal power  statistics.}
    \label{fig:cdf_powers}
\end{figure}

To further evaluate the effectiveness of the proposed learning model, we compare the \acp{CDF} of \ac{UL} and \ac{DL} powers obtained from the optimal closed-form solution~\cite{miretti2022closed} and those predicted by the trained model on the test set. As shown in \figurename~\ref{fig:cdf_powers}, the resulting curves are nearly overlapping in both cases, indicating that the model successfully learns the underlying power allocation strategy for the predicted clusters. This close agreement demonstrates that the predicted powers not only approximate the optimal allocation with high fidelity but also preserve the statistical distribution of transmit powers across \acp{UE}. Consequently, the proposed approach achieves reliable power prediction while avoiding the computational burden of solving the closed-form optimization problem. {\color{black}  This behavior is explained by the fact that the optimal power allocation is a mapping of large-scale channel statistics which highly depend on \ac{UE}-\ac{AP} geometry, and thus the trained model learns to approximate this mapping accurately even for unseen \ac{UE} deployments of the test set.}

\subsection{Robustness to input errors}

\begin{figure}[t!]
    \centering
    % \ac{UL} Subfigure
    \begin{subfigure}[b]{\columnwidth}
        \centering
        \includegraphics[width=\textwidth]{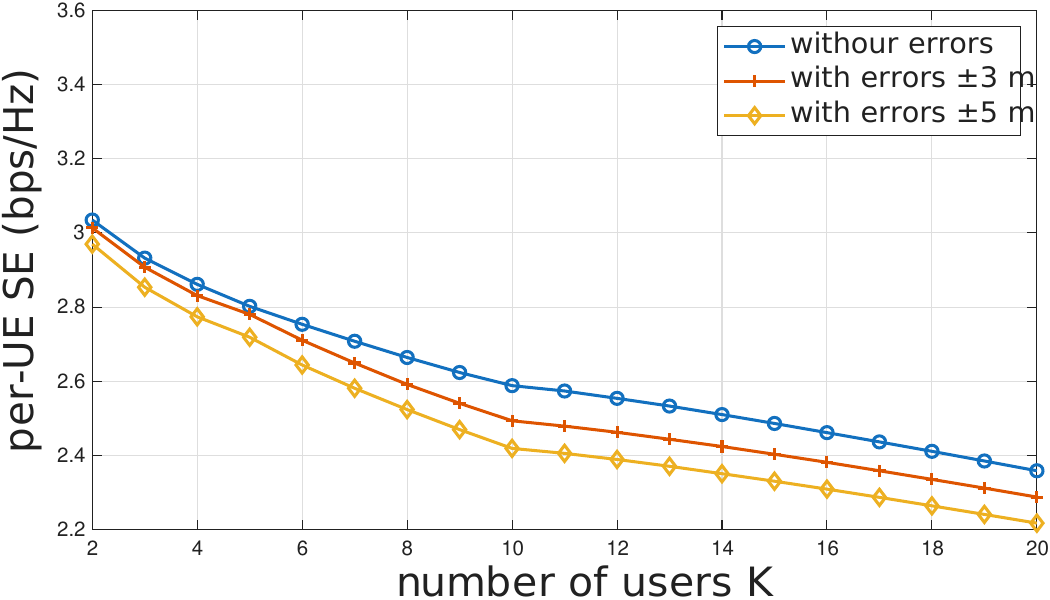} % Replace with your actual \ac{UL} figure filename
        \caption{\acs{UL}}
        \label{fig:ul_noise}
    \end{subfigure}
    \hfill
    % \ac{DL} Subfigure
    \begin{subfigure}[b]{\columnwidth}
        \centering
        \includegraphics[width=\textwidth]{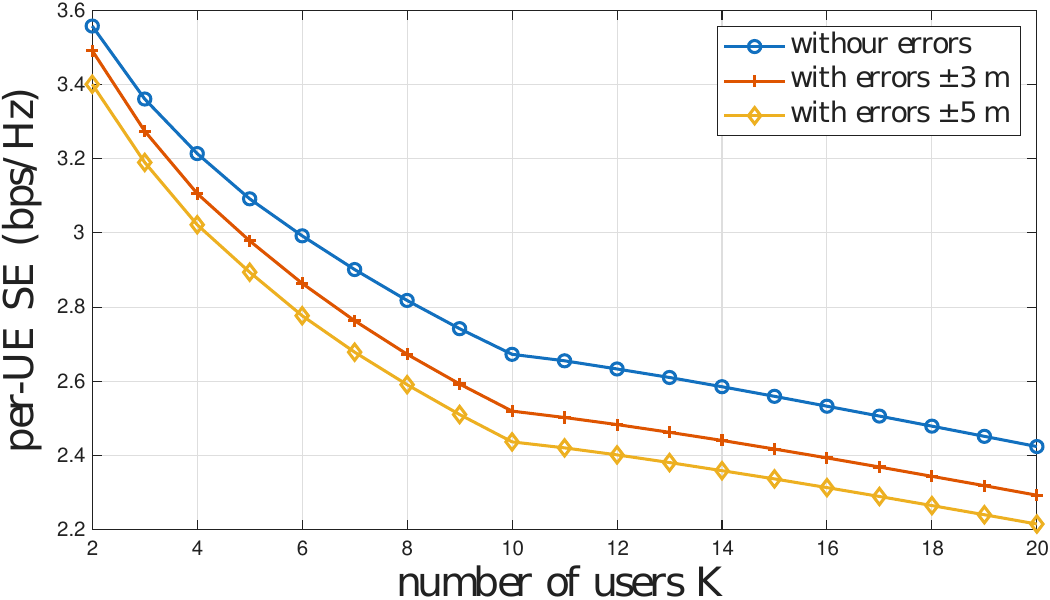} % Replace with your actual \ac{DL} figure filename
        \caption{\acs{DL}}
        \label{fig:dl_noise}
    \end{subfigure}
    \caption{Average per-\acs{UE} \acs{SE} for different \acs{UE} counts and input error levels. The trained model remains robust to input errors.}
    \label{fig:input_errors}
\end{figure}

In \figurename~\ref{fig:input_errors}, we analyze the robustness of the proposed learning model against input uncertainty. During training, as stated in \eqref{eq:trainingNoise}, the model is exposed to user position inputs corrupted by random Gaussian noise with a standard deviation of $1\meter$. To assess generalization and robustness, the model is evaluated under three test conditions: 
\begin{enumerate*}[label=\emph{\roman*})]
    \item input with a fixed standard deviation error of $3\meter$,
    \item input with a fixed standard deviation error of $5\meter$, and
    \item an ideal scenario with error-free input.
\end{enumerate*}

The results show that the per-\ac{UE} \ac{SE} decreases only slightly as the input error level increases, for both \ac{UL} and \ac{DL}. Specifically, the maximum gap between the error-free and $5\meter$ error cases is approximately $0.29\bpsHz$ in \ac{UL} and $0.24\bpsHz$ in \ac{DL} across all user counts, demonstrating the resilience of the proposed model. Moreover, the relative trends across different user loads are preserved, indicating that the model can generalize its learned spatial relationships even when the inputs are affected by localization errors, thus ensuring practical robustness for real-world deployments where user position estimates are inherently imperfect.

\subsection{Comparison with other Transformer variants}
\begin{table*}[htbp]
\centering
\caption{Performance comparison of Transformer variants in terms of complexity and efficiency metrics.}
\label{tab:model_comparison}
\begin{tabular}{@{}lcccccc@{}}
\toprule
\emph{\textbf{Model}} & \textbf{Parameters} & \textbf{Memory (MB)} & \textbf{\acsp{FLOP}} & \textbf{Latency (ms)} & \textbf{Training Time (s/epoch)} \\ \midrule
\emph{\ac{ELU}-CosFormer}   & $103{,}378$           & $0.39$                 & $1{,}056{,}820$   & $3.42$                  & $858.27 $                         \\
\emph{CosFormer}       & $103{,}378$           & $0.39$                 & $1{,}056{,}820$   & $3.40$                  & $915.82$                          \\
\emph{Transformer}     & $846{,}866$           & $3.23$                 & $5{,}682{,}260$   & $4.42$                  & $1447.69$                         \\
\emph{Longformer}      & $7{,}102{,}738$       & $27.09$                & $635{,}398{,}900$ & $680.99$                & $1866.63$                         \\
\emph{Performer}       & $103{,}378$           & $0.39$                 & $1{,}041{,}460$   & $4.40$                  & $1076.56$                         \\ \bottomrule
\end{tabular}
\end{table*}

\begin{figure}[t!]
    \centering
    \begin{subfigure}[b]{\columnwidth}
        \centering
        \includegraphics[width=\textwidth]{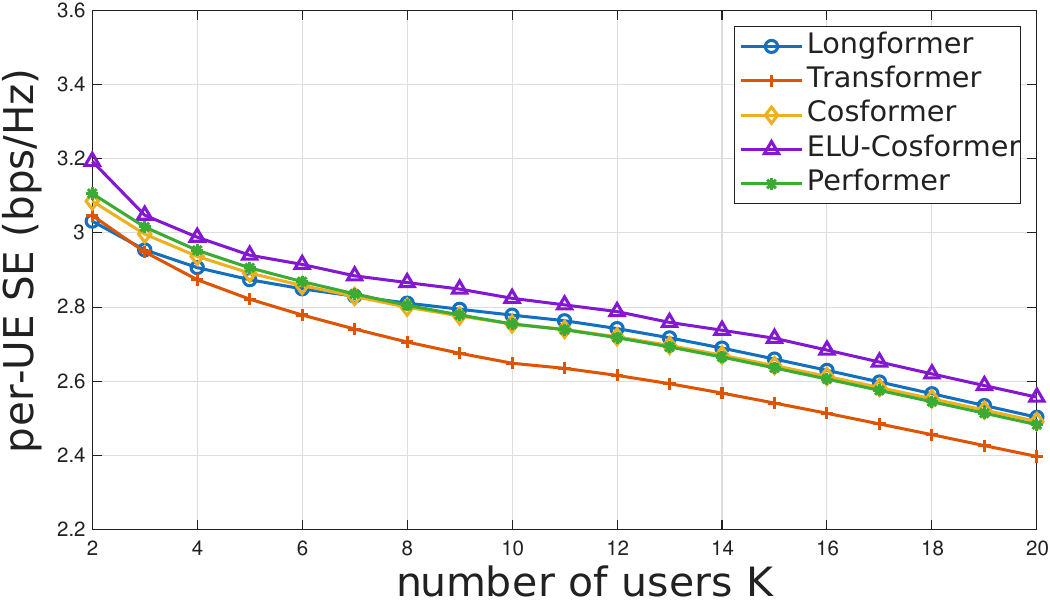}
        \caption{\acs{UL}}
        \label{fig:ul_se_comparison}
    \end{subfigure}
    \hfill
    \begin{subfigure}[b]{\columnwidth}
        \centering
        \includegraphics[width=\textwidth]{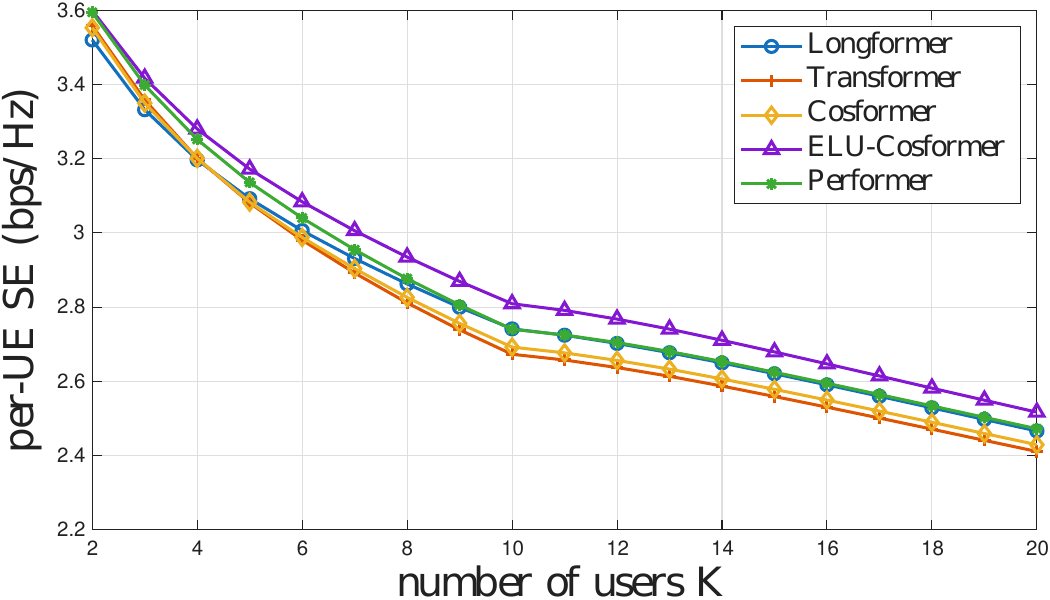}
        \caption{\ac{DL}}
        \label{fig:dl_se_comparison}
    \end{subfigure}
    \caption{Comparison of per-\acs{UE} \acs{SE} across different Transformer variants for \acs{UL} and \acs{DL} scenarios. Our modified CosFormer maintains higher \acs{SE}.}
    \label{fig:se_linear}
\end{figure}

\figurename~\ref{fig:se_linear} presents a comparative analysis of \ac{UL} and \ac{DL} per-\ac{UE} \ac{SE} achieved by the Transformer and four linear attention variants. Linformer is excluded because it cannot accommodate variable \ac{UE} counts without additional preprocessing or architectural modifications~\cite{wang2020linformer}.

{\color{black}In both \ac{UL} and \ac{DL} scenarios, \ac{ELU}-CosFormer consistently achieves the highest \ac{SE} across all user counts. This result highlights the benefit of integrating the \ac{ELU} activation into the CosFormer attention kernel. As discussed in \sectionname~\ref{sec:proposed}, the combination of \ac{ELU} with the cosine-based attention mechanism improves the model's ability to capture the spatial interactions between \acp{UE} and \acp{AP}. Consequently, \ac{ELU}-CosFormer produces more effective \ac{AP} clustering and power allocation decisions, maintaining strong performance even as the number of users increases under the considered fairness-driven objective.}

In addition to \ac{SE}, Table~\ref{tab:model_comparison} compares the models in terms of complexity and efficiency metrics. All models were evaluated under consistent conditions to ensure fair comparison, with latency and training time reported for $K=10$ \acp{UE}. 

\begin{itemize}
  \item \emph{Parameters}: Total number of trainable weights in the model.
  \item \emph{Memory (MB)}: Estimated memory footprint during inference, reflecting \ac{GPU} usage.
  \item \emph{\acsp{FLOP}}: Floating-point operations per forward pass, indicating computational complexity.
  \item \emph{Latency (ms)}: Average inference time, measuring real-time responsiveness.
  \item \emph{Training Time (s/epoch)}: Time per training epoch, capturing training efficiency.
\end{itemize}

\ac{ELU}-CosFormer and CosFormer share the same architecture and parameter count, differing only in their activation functions-\ac{ELU} and \ac{ReLU}, respectively. Although \ac{ELU} is computationally more intensive, \ac{ELU}-CosFormer achieves slightly lower training time per epoch, due to smoother gradient flow that enables more efficient backpropagation and faster convergence. In contrast, \ac{ReLU}’s simplicity gives CosFormer a marginal advantage in inference latency, as \ac{ReLU} operations are highly optimized for \ac{GPU} execution. Performer maintains a low parameter count and memory footprint comparable to CosFormer, but exhibits higher training and inference times due to the computational overhead of random feature projections and exponentiation in its kernel mapping \cite{choromanski2020rethinking}, which are less \ac{GPU}-efficient than CosFormer’s cosine reweighting.

The standard Transformer exhibits substantially higher \acsp{FLOP} and memory requirements, owing to its quadratic attention mechanism and larger parameter count, resulting in increased latency and longer training time. Longformer shows the highest latency and training time overall, reflecting its significantly larger parameter count relative to the other models.

In summary, the results in \figurename~\ref{fig:se_linear} and \tablename~\ref{tab:model_comparison} indicate that the proposed \ac{ELU}-CosFormer achieves the most balanced and scalable performance, validating its design for fairness-aware resource optimization.

\section{Computational Complexity Analysis}\label{sec:complexity}
When the number of users $K$ increases, the per-sample computational complexity of our model is dominated by the linearized attention term $\mathcal{O}(M \cdot d_{\text{mod}} \cdot K)$, which scales linearly with $K$. This linear scaling represents a clear advantage compared to prior Transformer-based approaches~\cite{kocharlakota2024pilot,chafaa2025transformer}, where the complexity grows quadratically as $\mathcal{O}(M \cdot d_{\text{mod}} \cdot K^2)$, and iterative optimization methods~\cite{farooq2020accelerated}, which scale as $\mathcal{O}(T \cdot L \cdot K^2)$ with $T$ iterations. It is also favorable compared to the optimal closed-form solution that require cubic scaling $\mathcal{O}(K^3)$, which quickly become prohibitive in dense networks.

For illustration, we benchmark the runtime on a \ac{CPU} for $K=40$, computing both \ac{AP}-\ac{UE} clusters and \ac{UL}/\ac{DL} powers using our predicted \ac{UC}-\ac{CF} scheme versus \ac{DCC} (with closed-form solution for power computation). The proposed \ac{UC}-\ac{CF} model requires only $9.2\ms$, whereas \ac{DCC} takes $31.6\second$. This orders-of-magnitude speedup, achieved through linearized attention, highlights the significant efficiency gain of our approach and makes it more suitable for large-scale deployments.

{\color{black}Beyond asymptotic complexity, several practical aspects reinforce the advantage of our model. The linear attention mechanism reduces memory usage compared to quadratic approaches, enabling scalability to larger user counts and deployment on resource-constrained edge devices. This efficiency also translates into lower energy consumption, aligning with sustainability goals in next-generation wireless networks. Moreover, the one-shot inference design avoids iterative overhead, ensuring real-time responsiveness to dynamic user distributions and mobility. Importantly, the model maintains near-optimal \ac{SE} despite its reduced complexity, confirming that efficiency does not compromise accuracy. Finally, linear attention benefits more from parallelization on   \acl{GPU}  and \acl{TPU} \cite{pal2019gpuparallel, jouppi2017tpu}, further widening the performance gap with quadratic and iterative methods.}

 %%%%%%%%%%%%%%%%%%%%%%%%%%%%%%%%%%%%%%%%%%%%%%%%%%%%%%%%%%%%%%
 %                                CONCLUSION
%%%%%%%%%%%%%%%%%%%%%%%%%%%%%%%%%%%%%%%%%%%%%%%%%%%%%%%%%%%%%%%%%%   
%\vspace{-0.35in}

 \section{Conclusion}\label{sec:conclusion}
{\color{black} This paper proposed a scalable and flexible deep learning model for joint \ac{AP} clustering and power allocation in \acs{UC} \acs{CF} \acs{mMIMO}  systems. The method follows a dynamic supervised learning paradigm in which optimal power  labels are generated on the fly for the predicted \ac{AP} clusters to train a CosFormer-based model that maps \ac{UE}-\ac{AP} spatial configurations to the corresponding resource allocation decisions. By relying primarily on spatial information and leveraging a customized  attention mechanism, the proposed framework avoids pilot contamination, adapts seamlessly to varying user loads, and enables resource allocation decisions without requiring instantaneous channel  information as input. This allows the joint clustering and power control to be decoupled from the data detection stage.

Numerical evaluations demonstrated that the proposed framework achieves near-optimal \ac{SE} while reducing computational complexity and signaling overhead. In particular, while the  complexity of the optimal solution scales as $\mathcal{O}(K^3)$ with the number of \acp{UE}, the proposed model performs resource allocation through a single  inference step and scales linearly with $K$. Furthermore, since the model relies only on spatial information during inference, it eliminates the need for exchanging instantaneous channels for resource  optimization, thereby reducing signaling overhead. In addition, the  clustering learned by the model limits the number of \acp{AP} serving each \ac{UE} to the most relevant ones, which further reduces coordination and signaling requirements between \acp{AP} and the \ac{CPU}.
}

Future research directions include investigating distributed learning strategies among \acp{AP} or clusters to further reduce centralization requirements. Another promising direction is the integration of temporal information, enabling the model to capture user mobility, traffic variations, and time-dependent channel dynamics. Extending the architecture with temporal attention mechanisms may enable proactive resource allocation and enhance robustness in highly dynamic operating conditions. {\color{black} Additionally, further validation could use ray-tracing datasets or real measurement campaigns. }

\bibliographystyle{IEEEtran}
\bibliography{biblio}

\end{document}